\documentclass[pra,a4paper,aps,twocolumn,superscriptaddress,longbibliography]{revtex4-1}
\usepackage{amsmath, empheq,braket,amsthm,amssymb,graphics,amsfonts,array,color,subfigure}
\usepackage{graphicx}
\definecolor{myurlcolor}{rgb}{0,0,0.7}
\definecolor{myurlcolor1}{rgb}{0,0.7,0.1}
\definecolor{myrefcolor}{rgb}{0,0,0.7}
\usepackage[unicode=true,pdfusetitle, bookmarks=true,bookmarksnumbered=false,bookmarksopen=false, breaklinks=false,pdfborder={0 0 0},backref=false,colorlinks=true, linkcolor=myrefcolor,citecolor=myurlcolor1,urlcolor=myurlcolor]{hyperref}
\usepackage{lipsum,MnSymbol}
\usepackage[utf8]{inputenc}  
\usepackage[T1]{fontenc}

\begin{document}

\title{Quantum Wigner entropy}

\author{Zacharie Van Herstraeten}
\email{zvherstr@ulb.ac.be}
\affiliation{Centre for Quantum Information and Communication, \'Ecole polytechnique de Bruxelles, CP 165/59, Universit\'e libre de Bruxelles, 1050 Brussels, Belgium}
\author{Nicolas J. Cerf}
\email{ncerf@ulb.ac.be}
\affiliation{Centre for Quantum Information and Communication, \'Ecole polytechnique de Bruxelles, CP 165/59, Universit\'e libre de Bruxelles, 1050 Brussels, Belgium}

\begin{abstract}
We define the Wigner entropy of a quantum state as the differential Shannon entropy of the Wigner function of the state. 
This quantity is properly defined only for states that possess a positive Wigner function, which we name Wigner-positive states, but we argue that it is a proper measure of quantum uncertainty in phase space. 
It is invariant under symplectic transformations (displacements, rotations, and squeezing) and we conjecture that it is lower bounded by $\ln\pi +1$ within the convex set of Wigner-positive states. 
It reaches this lower bound for Gaussian pure states, which are natural minimum-uncertainty states. 
This conjecture bears a resemblance with the Wehrl-Lieb conjecture, and we prove it over the subset of passive states of the harmonic oscillator which are of particular relevance in quantum thermodynamics. 
Along the way, we present a simple technique to build a broad class of Wigner-positive states exploiting an optical beam splitter and reveal an unexpectedly simple convex decomposition of extremal passive states. 
The Wigner entropy is anticipated to be a significant physical quantity, for example, in quantum optics where it allows us to establish a Wigner entropy-power inequality. It also opens a way towards stronger entropic uncertainty relations. 
Finally, we define the Wigner-Rényi entropy of Wigner-positive states and conjecture an extended lower bound that is reached for Gaussian pure states.
\end{abstract}

\maketitle

\section{Introduction}


The phase-space formulation of quantum mechanics provides a complete framework that echoes classical statistical mechanics.
Quantum states and quantum operators are described within this formulation by continuous functions of the pair of canonical variables $x$ and $p$.
These variables traditionally refer to the position and momentum observables, but are also isomorphic to the conjugate quadrature components of a mode of the electromagnetic field (we use this quantum optics nomenclature in the present paper). The conversion from quantum operators to quantum phase-space distributions is carried out via the Wigner-Weyl transform \cite{Case2008}, which maps any linear operator $\hat{A}$ into a distribution $A(x,p)$ as
\begin{equation}
	A(x,p)=
	\dfrac{1}{\pi\hbar}
	\int
	\exp\left(2ipy/\hbar\right)
	\bra{x-y}
	\hat{A}
	\ket{x+y}
	\mathrm{d}y  ,
\end{equation}
where $\hbar$ denotes the Planck constant (we set $\hbar =1$ in the remainder of this paper). Accordingly, the Wigner function of a quantum state is the Wigner-Weyl transform of its density operator $\hat{\rho}$, written as $W(x,p)$. The Wigner function comes as close to a probability distribution in phase space as allowed by quantum mechanics. It indeed shares most properties of a classical probability distribution. Notably, the marginal distributions of $W(x,p)$ coincide with the probability distributions for $x$ and $p$, respectively $\rho_x(x)=\bra{x}\hat{\rho}\ket{x}$ and $\rho_p(p)=\bra{p}\hat{\rho}\ket{p}$, as it can easily be shown that $\int W(x,p)\, \mathrm{d}p=\rho_x(x)$ and $\int W(x,p)\, \mathrm{d}x=\rho_p(p)$.
Also, the expectation value of any operator $\hat{A}$ in state $\hat{\rho}$ is straightforwardly computed from its Wigner function through the overlap formula \cite{Leonhardt2010}:
\begin{equation}
	\langle \hat{A}\rangle
	=
	\mathrm{Tr}\left[\hat{A}\, \hat{\rho}\right]
	=
	2\pi\iint A(x,p)\, W(x,p)\, \mathrm{d}x \, \mathrm{d}p.
	\label{eq:overlap_formula}
\end{equation}
However, it is well known that the Wigner function is not a true probability distribution as it lacks positiveness  \cite{Zyczkowski2004}. For example, all pure non-Gaussian states have a Wigner function that admits negative regions as a consequence of the Hudson theorem \cite{Hudson1974}. This is the price to pay to the Heisenberg uncertainty principle, which forbids the joint definition of noncommuting variables $x$ and $p$. Hence, several common functionals of probability distributions, such as the Shannon differential entropy, become in general ill defined if applied to Wigner functions.

In contrast, there exists a well-known distribution in quantum phase space that behaves as a genuine probability distribution, namely the Husimi $Q$ function  \cite{Leonhardt2010}, defined as $Q(\alpha)= \bra{\alpha}\hat{\rho}\ket{\alpha} /\pi$. 
It corresponds to the probability to measure state $\hat{\rho}$ in a coherent state $\ket{\alpha}$.
Remember that a coherent state $\ket{\alpha}$ is an eigenstate of the annihilation operator $\hat{a}=\left(\hat{x}+i\hat{p}\right)/\sqrt{2}$ with eigenvalue $\alpha$. 
Splitting the complex parameter $\alpha$ into two real parameters $x$ and $p$ such that $\alpha=x+ip$ gives
\begin{equation}
	Q(x,p) =  \dfrac{1}{\pi}   
	\bra{x+ip}\hat{\rho}\ket{x+ip} \,  .
	\label{eq:def_husimi}
\end{equation}
Despite lacking the nice properties of the Wigner function such as the overlap formula \eqref{eq:overlap_formula}, the Husimi function has the advantage of being positive, hence it admits a properly defined entropy. The Shannon differential entropy of the Husimi function is indeed known as the Wehrl entropy and is defined as $h\left(Q\right)=-\iint Q(x,p)\ln Q(x,p)\, \mathrm{d}x \, \mathrm{d}p$.
This entropy is at the core of the Wehrl conjecture \cite{Wehrl1979}, later proven by Lieb \cite{Lieb1978,Lieb2002}, which states that the Wehrl entropy is lower-bounded by $\ln\pi+1$ and that the only minimizers of $h(Q)$ are the coherent states \footnote{The Wehrl conjecture is also often written as $h\left(Q\right)\ge 1$, which simply originates from a different convention. If the Husimi Q-function is defined without the $1/\pi$ prefactor but, instead, this additional prefactor is inserted in the definition of $h\left(Q\right)$, we get an additive constant $-\ln\pi$ which shifts the lower bound to 1.}.

\begin{figure}
	\includegraphics[width=5cm]{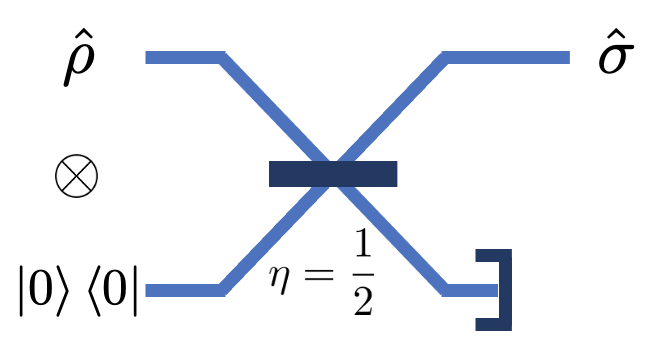}
	\caption{
		Reduced output state $\hat{\sigma}$ of a balanced beam splitter (of transmittance $\eta=1/2$) when the input state is $\hat{\rho}$, as described in Eq. \eqref{eq:sigma_output_bbs_vacuum}.
		The Wigner function of $\hat{\sigma}$ coincides with the Husimi $Q$ function of $\hat{\rho}$; hence it is positive. Consequently, the Wigner entropy of $\hat{\sigma}$ is equal to the Wehrl entropy of $\hat{\rho}$.}
	\label{fig:bbs_vacuum}
\end{figure}

Interestingly, there is a link between the Husimi function and Wigner function of a state, as can simply be understood using the quantum optics language. 
To this purpose, recall that the vacuum state $\ket{0}$ [or ground state of the harmonic oscillator $\hat{H} = \left(\hat{p}^2+\hat{x}^2\right)/2$ in natural units] admits the Wigner function $W_0\left(x,p\right)=\exp\left(-x^2-p^2\right)/\pi$. 
Since a coherent state  $\ket{\alpha}$ is a displaced vacuum state, its Wigner function is then $W_\alpha(x,p)=W_0\left(x',p'\right)$, where $x'=x-\sqrt{2}\, \mathrm{Re}(\alpha)$ and $p'=p-\sqrt{2}\, \mathrm{Im}(\alpha)$.
Using this, the Husimi function can be expressed from the overlap formula \eqref{eq:overlap_formula} as:
\begin{eqnarray}
	\hspace{-0.5cm} Q(x,p) &=& \dfrac{1}{\pi}   \mathrm{Tr}\left[ \ket{x+ip}\!\bra{x+ip} \, \hat{\rho}\right]  \nonumber \\
	&=& 2\iint
	W_0\left(\tilde{x}-\sqrt{2}x,\tilde{p}-\sqrt{2}p\right)
	W(\tilde{x},\tilde{p}) \,
	\mathrm{d}\tilde{x} \, 
	\mathrm{d}\tilde{p}.   
\end{eqnarray} 
Thus, it appears that $Q$ is a convolution between $W$ and $W_0$,  with a rescaling factor of $\sqrt{2}$.
In the language of random variables (and provided $W$ is non-negative), we could say that if $(\tilde{x},\tilde{p})$ is distributed according to $W$ and $(x_0,p_0)$ is distributed according to $W_0$, then $(x,p)$ is distributed according to $Q$, with 
\begin{equation}
x = \left(\tilde{x}-x_0\right)/\sqrt{2} \quad \textrm{and} \quad p = \left(\tilde{p}-p_0\right)/\sqrt{2} .
\end{equation}
This is a familiar relation in quantum optics, describing the action of a beam splitter of transmittance $\eta=1/2$ onto the state $\hat{\rho}$ and the vacuum state. Defining $\hat{\sigma}$ as the reduced state of the corresponding output of the beam splitter, as shown in Fig. \ref{fig:bbs_vacuum}, 
we conclude that the Wigner function of $\hat{\sigma}$ is precisely the Husimi function of $\hat{\rho}$, namely,
\begin{equation}
W_{\hat{\sigma}}(x,p) =Q_{\hat{\rho}}(x,p),
\label{eq:fundamental}
\end{equation}
where
\begin{equation}
\hat{\sigma}=
\mathrm{Tr}_2\left[\hat{U}_{\frac{1}{2}}\left(\hat{\rho}\otimes\ket{0}\bra{0}\right)\hat{U}_\frac{1}{2}^\dagger\right] .
\label{eq:sigma_output_bbs_vacuum}
\end{equation}
Here $\hat{U}_{\frac{1}{2}}$ denotes the beam-splitter unitary of transmittance $\eta=1/2$, while $\mathrm{Tr}_2$ denotes a reduced trace over one of the modes, say the second mode. 
From Eq. \eqref{eq:fundamental}, it appears that the entropy of the Wigner function of $\hat{\sigma}$ is nothing else but the Wehrl entropy of $\hat{\rho}$ in this particular setup.
A natural question then arises : can we give an intrinsic meaning to the entropy of a Wigner function independently of this particular setup?

In this paper, we will answer by the affirmative. First, let us notice that the setup of Fig. \ref{fig:bbs_vacuum} ensures that the output state $\hat{\sigma}$ always has a positive Wigner function (see Appendix \ref{apd:bbs_wigner_positive}). In general, we will denote the quantum states admitting a positive Wigner function [i.e., states such that $W(x,p)\ge 0$, $\forall x,p$] as \textit{Wigner-positive} states. For such states, it is possible to compute the Shannon differential entropy of their Wigner function.  We make the leap and define the \textit{Wigner entropy} of any Wigner-positive state $\hat{\rho}$ as 
\begin{equation}
	h\left(W\right)   =  -\iint  W(x,p)  \,  \ln  W(x,p) \,   \mathrm{d}x \, \mathrm{d}p
\label{eq-def-Wigner-entropy}	
\end{equation}
where 
\begin{equation}
W(x,p) = \frac{1}{\pi} \int \exp\left(2ipy\right) \bra{x-y} \hat{\rho} \ket{x+y} \mathrm{d}y
\label{eq-def-Wigner-function}
\end{equation}
is the Wigner function of  $\hat{\rho}$. We argue that, although it is limited to Wigner-positive states, the Wigner entropy is a natural measure in order to characterize quantum uncertainty in phase space: it bears information about the uncertainty of the marginal distributions of the $x$ and $p$ variables as well as their correlations in phase space. In contrast with the Wehrl entropy, it is not the classical entropy of the outcome of a specific measurement, namely, a joint $(x,p)$ measurement (called heterodyne detection or eight-port homodyne detection in quantum optics). Of course, in the special case where a Wigner-positive state can be prepared using the setup of Fig. \ref{fig:bbs_vacuum}, its Wigner entropy can be viewed simply as the Wehrl entropy of the corresponding input state, but the definition goes further and the Wigner entropy remains relevant for Wigner-positive states that \textit{cannot} be built in this way.

The Wigner entropy $h(W)$ enjoys interesting properties. First, unlike the Wehrl entropy $h(Q)$, it is invariant under symplectic transformations (displacement, rotation, and squeezing) in phase space. Such transformations, which are ubiquitous in quantum optics, correspond to the set of all Gaussian unitaries in state space. We stress that a sensible measure of phase-space uncertainty must remain invariant under symplectic transformations since these are also area-preserving transformations in phase space. In contrast, $h(Q)$ is greater for squeezed states than for coherent states. As it can be understood from Fig. \ref{fig:bbs_vacuum}, this preference simply originates from the fact that one input of the balanced beam splitter is itself a coherent state. Second, the Wigner entropy $h(W)$ can be related to the entropy of the marginal distributions $h\left(\rho_x\right)$ and $h\left(\rho_p\right)$, but also encompasses the $x$-$p$ correlations. Shannon information theory establishes a relation between the entropy of a joint distribution and its marginal entropies, namely, $h(x,p)=h(x)+h(p)-I$, where $I\ge 0$ is the mutual information \cite{Cover1991}. Applied to the Wigner entropy, this gives the inequality $h(W)\leq h\left(\rho_x\right)+h\left(\rho_p\right)$.
This means that a lower bound on the Wigner entropy implies in turn a lower bound on the sum of the marginal entropies.

In the light of these considerations, we introduce a conjecture on the Wigner entropy, which resembles the Wehrl conjecture. As anticipated in \cite{Hertz2017}, we conjecture that the Wigner entropy of any Wigner-positive state $\hat{\rho}$ satisfies
\begin{equation}
	h\left(W\right)\geq\ln\pi+1.
	\label{eq:wig_conj}
\end{equation}
As we will show, this bound is reached by all Gaussian pure states, which appears consistent with the Hudson theorem \cite{Hudson1974}.
It implies (but is stronger than) the entropic uncertainty relation of Białynicki-Birula and Mycielski \cite{Bialynicki1975}, namely, $h\left(\rho_x\right)+h\left(\rho_p\right)\geq\ln\pi+1$.
Importantly, conjecture \eqref{eq:wig_conj} also implies the Wehrl conjecture since we have shown that the Husimi function of any state $\hat {\rho}$ is the Wigner function of some Wigner-positive state $\hat {\sigma}$ in a particular setup (see Fig. \ref{fig:bbs_vacuum}). However, the converse is not true as there exist Wigner-positive states whose Wigner function cannot be written as the Husimi function of a physical state (an example will be shown in Sec. \ref{sec:results}).

The paper is organized as follows. In Sec. \ref{sec:wig_entropy}, we start by recalling some basics of the symplectic formalism and then define the Wigner entropy of a Wigner-positive state as a distinctive information-theoretical measure of its uncertainty in phase space. In Sec. \ref{sec:wig_pos}, we discuss the characterization of the set of Wigner-positive states and focus on the particular subset of phase-invariant Wigner-positive states. Then, in  Sec. \ref{sec:results}, we turn to the main conjecture and provide a proof for some special case of phase-invariant Wigner-positive states, namely the passive states. Finally, we conclude in Sec. \ref{sec:conclusion} and provide an example application of the Wigner entropy, namely the Wigner entropy-power inequality. Further, in Appendix~\ref{sect-wigner-renyi}, we extend the Wigner entropy and define the Wigner-R\'enyi entropy of Wigner-positive states. We also discuss a natural extension of the conjectured lower bound. In Appendix~\ref{apd:bbs_wigner_positive}, we present a quantum-optics-inspired method for generating a large variety of Wigner-positive states with a balanced beam splitter, extending on Fig.~\ref{fig:bbs_vacuum}. Appendix~\ref{apd:mixture_2phot} is devoted to the detailed analysis of the set of Wigner-positive states when considering the Fock space restricted to two photons as this provides a helpful illustration of our results. Finally, Appendix~\ref{apd:formula_extremal_states} provides more details on the derivation of the formula [Eq. \eqref{eq:extremal_states_formula}] at the heart of our proof.

\section{Wigner entropy of a state}
\label{sec:wig_entropy}

In this paper, we restrict our considerations to a single bosonic mode (one harmonic oscillator) for simplicity, although the definition of the Wigner entropy and the corresponding conjecture should extend to the multidimensional case. Let us briefly review the symplectic formalism for one bosonic mode. Let $\mathbf{\hat{x}} = (\hat{x}, \hat{p})^\intercal$ be the vector of quadrature operators (or position and momentum canonical operators) satisfying $[\hat{x}_j,\hat{x}_k]= i \, \Omega_{jk}$, with the matrix
\begin{equation}
\mathbf{\Omega} =\begin{pmatrix}
		0 & 1
		\\
		-1 & 0
	\end{pmatrix}
\end{equation}
being the symplectic form. The coherence vector (also called the displacement vector) of a state $\hat{\rho}$ is defined as  
\begin{equation}
\mathbf{c} = \langle  \mathbf{\hat{x}}  \rangle  \coloneq   \mathrm{Tr}( \mathbf{\hat{x}}  \,  \hat{\rho}),
\end{equation}
where $\langle  \cdot \rangle$ stands for the expectation value in state $\hat{\rho}$, while the covariance matrix $\mathbf{\Gamma}$ of state $\hat{\rho}$ is defined as
\begin{equation}
\Gamma_{jk}= \langle \{ \hat{x}_j -  \langle \hat{x}_j \rangle , \hat{x}_k - \langle \hat{x}_k \rangle \} \rangle
\end{equation}
where $\{ \cdot,\cdot \}$ stands for the anticommutator.
The set of Gaussian states contains those for which the Wigner function $W(x,p)$ is Gaussian; hence these states are completely characterized by their first- and second-order moments  $\mathbf{c}$ and  $\mathbf{\Gamma}$. The set of Gaussian unitaries in state space is isomorphic to the set of symplectic transformations in phase space. Formally, a symplectic transformation is an affine map on the space of quadrature operators which is defined by a symplectic matrix $\mathbf{S}$ and a displacement vector $\mathbf{d}$, namely,
\begin{equation} 
\mathbf{\hat{x}}  \to  \mathbf{S}\mathbf{\hat{x}}+\mathbf{d}.
\end{equation}
The symplectic matrix $\mathbf{S}$ is a real matrix that must preserve the symplectic form, that is, $\mathbf{S} \mathbf{\Omega} \mathbf{S}^\intercal = \mathbf{\Omega}$, which implies in particular that $\det\mathbf{S}=1$. The displacement vector $\mathbf{d}$ is an arbitrary real vector. The first- and second-order moments of a state $\rho$ evolve under such a symplectic transformation as
\begin{equation} 
\mathbf{c}  \to  \mathbf{S} \mathbf{c} +\mathbf{d}  \, ,  \qquad  \mathbf{\Gamma}=\mathbf{S} \mathbf{\Gamma} \mathbf{S}^\intercal  \, .
\end{equation}
In the special case of Gaussian states, this completely characterizes the evolution of the state under the Gaussian unitary.

The core of this paper is the definition of an information-theoretical measure of uncertainty in phase space, which we call the Wigner entropy $h(W)$, where $h(\cdot)$ denotes the Shannon differential entropy functional and $W(x,p)$ is the Wigner function of $\hat{\rho}$ [see Eqs. \eqref{eq-def-Wigner-entropy} and \eqref{eq-def-Wigner-function}]. As already mentioned, it only applies to Wigner-positive states since, otherwise, the definition of the entropy entails the logarithm of a negative number. We note it as a functional of $W$ but, of course, it is eventually a functional of the state  $\hat{\rho}$ since $W$ itself depends on $\hat{\rho}$.

In contrast with the Shannon entropy of a discrete variable, the Shannon differential entropy of a continuous variable does not have an absolute meaning (it depends on the scale of the variable) and it becomes negative if the probability distribution is highly peaked \cite{Cover1991}. However, when applied to a Wigner function, a natural scale is provided here by the area $\hbar$ of a unit cell in phase space.  Hence, the Wigner entropy has a meaning \textit{per se} and it is legitimate to conjecture a lower bound, namely, Eq.~\eqref{eq:wig_conj}, when setting $\hbar=1$. Further, it is natural to extend on this and consider a lower bound on the differential R\'enyi entropy of the Wigner function of any Wigner-positive state, a quantity that we define as the Wigner-R\'enyi entropy (see Appendix \ref{sect-wigner-renyi}).

The Wigner entropy $h(W)$ has the nice property to be invariant under symplectic transformations. Consider the symplectic transformation $\mathbf{\hat{x}}  \to \mathbf{\hat{x}'} =  \mathbf{S}\mathbf{\hat{x}}+\mathbf{d}$ and let us denote as $W$ and $W'$ the Wigner function of the input and output states, respectively. The change of variables corresponding to this transformation gives
\begin{equation}
W'(x',p') = \frac{ W(x,p)} {|\det\mathbf{S}|},
\end{equation}
which indeed implies that
\begin{eqnarray}
h(W') &=& -\iint   W'(x',p') \, \ln  W'(x',p') \, \mathrm{d}x' \, \mathrm{d}p'    \nonumber \\
&=& -\iint   W(x,p)  \ln  \left( \frac{ W(x,p)} {|\det\mathbf{S}|} \right) \, \mathrm{d}x \, \mathrm{d}p \nonumber \\
&=& h(W) +  \ln |\det\mathbf{S}| \nonumber \\
&=& h(W),
\end{eqnarray}
where we have used the fact that $W$ is normalized and the fact that $\mathbf{S}$ is a symplectic matrix ($\det\mathbf{S}=1$). 

Note that this invariance can also be understood as a sole consequence of the fact that symplectic transformations conserve areas in phase space since $\det\mathbf{S}=1$. Indeed, for any functional $F$, we have
\begin{eqnarray}
\lefteqn{    \iint   F\big( W'(x',p') \big) \, \mathrm{d}x' \, \mathrm{d}p'    }   \hspace{1cm} \nonumber \\
&&= \iint  F  \left( \frac{ W(x,p)} {|\det\mathbf{S}|} \right) \, |\det\mathbf{S}| \, \mathrm{d}x \, \mathrm{d}p \nonumber \\
&&= \iint   F\big( W(x,p) \big) \, \mathrm{d}x \, \mathrm{d}p.   
\end{eqnarray}

The special case of Gaussian states is very easy to deal with. A straightforward calculation shows that the Wigner entropy of a Gaussian state $\hat{\rho}$ is given by
\begin{equation}
h(W)= \ln \left( 2 \pi \sqrt{\det\mathbf{\Gamma}} \right) + 1 = \ln ( \pi / \mu )  + 1,
\end{equation}
where $\mu=\mathrm{Tr} \hat{\rho}^2 = 1/ (2  \sqrt{\det\mathbf{\Gamma}}) \le 1$ stands for the purity of the state.
All Gaussian states that are connected with a symplectic transformation obviously conserve their purity since $\det\mathbf{\Gamma'} = \det (\mathbf{S} \mathbf{\Gamma} \mathbf{S}^\intercal) = \det\mathbf{\Gamma}$, which confirms that their Wigner entropy is invariant. The lowest value of $h(W)$ among Gaussian states is then reached for pure states ($\mu=1$) and is given by $\ln \pi + 1$, as expected. This is the value of the Wigner entropy of all coherent states and squeezed states (regardless the squeezing parameter, squeezing orientation, and coherence vector). Accordingly, the Gaussian pure states would be the minimum-Wigner-uncertainty states. The difficult task remains, however, to prove that non-Gaussian Wigner-positive states cannot violate this lower bound (see Sec. \ref{sec:results}).

Provided this conjecture is valid, the Wigner function of any Wigner-positive state can be classically simulated from the Wigner function of the vacuum state (or any other Gaussian pure state). More precisely, information theory tells us that the difference $\Delta = h(W)- \ln \pi - 1$ can be viewed as the number of independent equiprobable random bits that are needed, on average, to generate deterministically one random $(x,p)$ instance drawn from the Wigner function of state $\rho$ from one random $(x,p)$ instance drawn from the Wigner function of the vacuum state (or any Gaussian pure state). Of course, this results holds at the asymptotic limit only, that is, around $N\times \Delta$ bits of extra randomness are needed for converting $N$ random instances of $(x,p)\sim W_0$ into $N$ random instances of $(x,p)\sim W$ by deterministic means when $N\to \infty$.

\section{Wigner-positive states}
\label{sec:wig_pos}

As explained in Sec. \ref{sec:wig_entropy}, the Wigner entropy naturally appears as an information-theoretic measure of uncertainty in phase space, but is only properly defined for positive Wigner functions. For this reason, we devote this section to the quantum states with positive Wigner functions, which we call \textit{Wigner-positive} states. Note that Wigner positivity is a particular case of $\eta$-positivity for $\eta=0$ \cite{Narcowich1989, Brocker1995}.
Quantum Wigner-positive states of a single mode are described by a Wigner function $W(x,p)$ that respects the condition
\begin{equation}
	W(x,p)\geq 0
	\qquad
	\forall x,p.
\end{equation}
Restricting to pure states, the set of Wigner-positive states is well known: the Hudson theorem establishes that Gaussian pure states are the only pure quantum states with a positive Wigner function \cite{Hudson1974}. When it comes to mixed states, however, the situation becomes more difficult since the mixing of states enables one to build non-Gaussian Wigner-positive states. The characterization of the set of Wigner-positive mixed states has been attempted \cite{Brocker1995,Mandilara2009}, but the resulting picture is somehow complex.
Just like writing a necessary and sufficient condition for a Wigner function to correspond to a positive-semidefinite density operator is a hard task, it appears cumbersome to express a necessary and sufficient condition for a density operator to be associated with a positive Wigner function.

\begin{figure}
	\includegraphics[width=3cm]{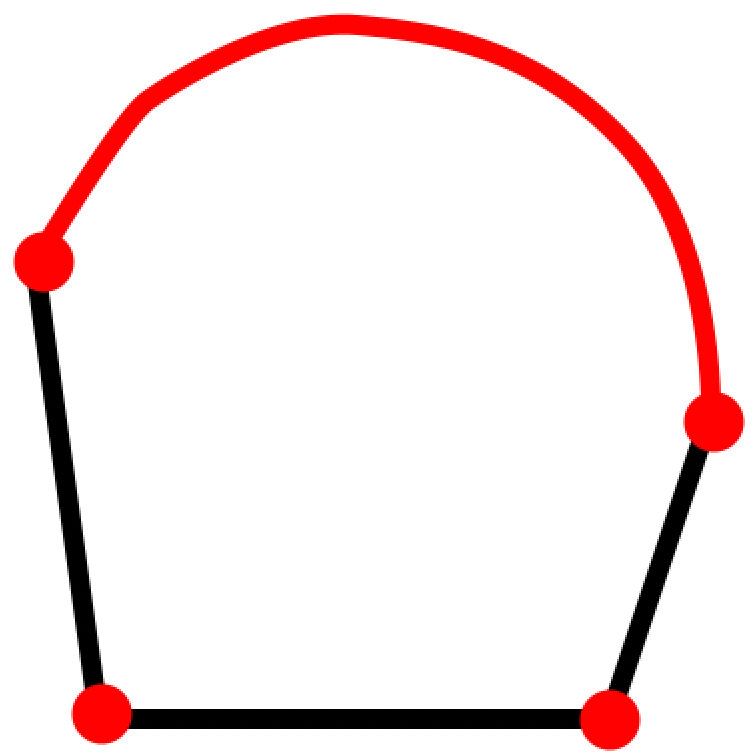}
	\caption{Schematic view of a convex set. The black and red points form all together the boundary of the convex set, while the red points are the extremal points on this boundary (note the existence of isolated extremal points as well as of a continuum of extremal points).}
	\label{fig:convex_set_extremal_points}
\end{figure}

On a more positive note, the set of Wigner-positive states is convex since a mixture of Wigner-positive states is itself Wigner positive. Taking advantage of this property, we may focus on the extremal states of the convex set, as pictured in Fig. \ref{fig:convex_set_extremal_points}. These are the states that cannot be obtained as a mixture of other states of the set. Conversely, any state of the set can be generated as a mixture of these extremal states. This brings a simplification in the proof of the main conjecture, namely, expressing a lower bound on the Wigner entropy of an arbitrary Wigner-positive state (see Sec. \ref{sec:results}). Indeed, the Shannon entropy being concave, the entropy of a mixture is lower-bounded by the entropy of its components, that is,
\begin{equation}
h(p_1 W_1+ p_2 W_2)\geq p_1 \, h(W_1) + p_2 \, h(W_2),
\end{equation}
where $p_1$ and $p_2$ are positive reals such that $p_1+p_2=1$. Hence, it is sufficient to prove the lower bound on $h(W)$ for the extremal Wigner-positive states in order to have a proof over the full set.

\subsection*{Several classes of  Wigner-positive states}

To be more specific, we define several sets of Wigner-positive quantum states, which  will be useful in the rest of this paper. As we will see, conjecture \eqref{eq:wig_conj} is trivially verified for some of them, while it remains hard to prove for others. 


\begin{itemize}
	\item $\mathcal{Q}$ :
	\textit{Physical quantum states}
	\\
	It is the convex set of all single-mode quantum states. 
	Their density operator $\hat{\rho}$ satisfies the three physicality conditions: Hermiticity, positive semidefiniteness and unit trace.
	Of course, they can have partly negative Wigner functions.
	
	\item $\mathcal{Q}_+$ :
	\textit{Wigner-positive quantum states}
	\\
	It is the subset of states in $\mathcal{Q}$ that have positive Wigner functions. 
	It is a convex set. All states within this set have a well-defined Wigner entropy and are the subject of conjecture \eqref{eq:wig_conj}.
	
	\item $\mathcal{G}$ :
	\textit{Gaussian states}
	\\
	It is the subset of states in $\mathcal{Q}_+$ that have a Gaussian Wigner function. It does not form a convex set since the mixture of Gaussian states does not need to be Gaussian, so we refer to its convex hull as $\mathcal{G}_c$.
	
	\item $\mathcal{C}$ :
	\textit{Classical states}
	\\
	According to Glauber's definition, classical states are mixtures of coherent states.
	They are characterized by a positive Glauber-Sudarshan $P$ function.
	By definition, $\mathcal{C}$ is a convex set and $\mathcal{C}\subset\mathcal{G}_c$ since coherent states are Gaussian states.
	
\end{itemize}

The extremal states of $\mathcal{C}$ and $\mathcal{G}_c$ are respectively coherent states and Gaussian pure states. 
For these two sets, conjecture \eqref{eq:wig_conj} is trivially verified since the Wigner entropy of Gaussian pure states is precisely $\ln\pi+1$ and since the entropy is concave.
Unfortunately, the convex closure of Gaussian states $\mathcal{G}_c$ is yet but a small fraction of the set of Wigner-positive states $\mathcal{Q}_+$.
As an evidence of this, we construct a wider set of Wigner-positive states by exploiting a technique relying on a balanced beam splitter (hence, we name this set as $\mathcal{B}$).

\begin{figure}
	\includegraphics[width=5cm]{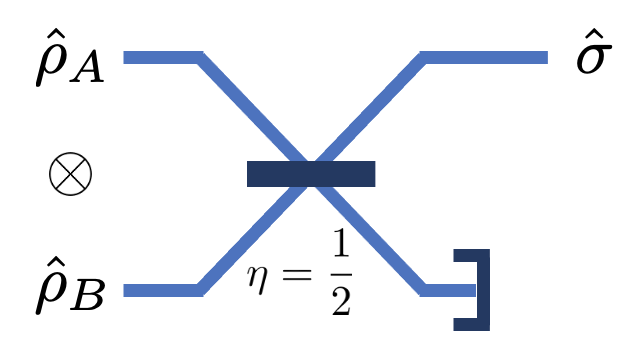}
	\caption{Beam-splitter state ${\hat\sigma}$ obtained at the output of a balanced beam splitter (of transmittance $\eta=1/2$). If the input is an arbitrary product state $\hat{\rho}_A\otimes\hat{\rho}_B$, then the reduced state of the output ${\hat\sigma}$ is guaranteed to be Wigner positive. This generalizes Fig. \ref{fig:bbs_vacuum}, where $\hat{\rho}_B=\ket{0}\!\bra{0}$. The set of beam-splitter states is denoted as $\mathcal{B}$. The convex hull of these states, denoted as $\mathcal{B}_c$, is obtained by sending any separable state (i.e., a mixture of product states) into a balanced beam splitter and tracing over one of the output modes. The whole set $\mathcal{B}_c$ is strictly included in the set of Wigner-positive states $\mathcal{Q}_+$. }
	\label{fig:bbs_schema}
\end{figure}

\begin{itemize}
	\item $\mathcal{B}$ :
	\textit{Beam-splitter states}
	\\
These are the states $\hat{\sigma}$ resulting from the setup depicted in Fig. \ref{fig:bbs_schema}. More precisely, a beam-splitter state $\hat{\sigma}$ denotes the reduced output state of a beam splitter with transmittance $\eta=1/2$ fed by a tensor product of two arbitrary states $\hat{\rho}_A$ and $\hat{\rho}_B$,
\begin{equation}
	\hat{\sigma} = 
	\mathrm{Tr}_2\left[
	\hat{U}_{\frac{1}{2}}
	\left(
	\hat{\rho}_A\otimes\hat{\rho}_B
	\right)
	\hat{U}_{\frac{1}{2}}^{\dagger}
	\right]  .
	\label{eq:bbs_wig_positive}
\end{equation}
We show in Appendix \ref{apd:bbs_wigner_positive} that state $\hat{\sigma}$ always possesses a positive Wigner function, regardless of $\hat{\rho}_A$ and $\hat{\rho}_B$. The sole condition is that the input state is a tensor product  and the beam splitter is balanced ($\eta=1/2$).
	
\end{itemize}

\begin{figure}
	\includegraphics[width=7cm]{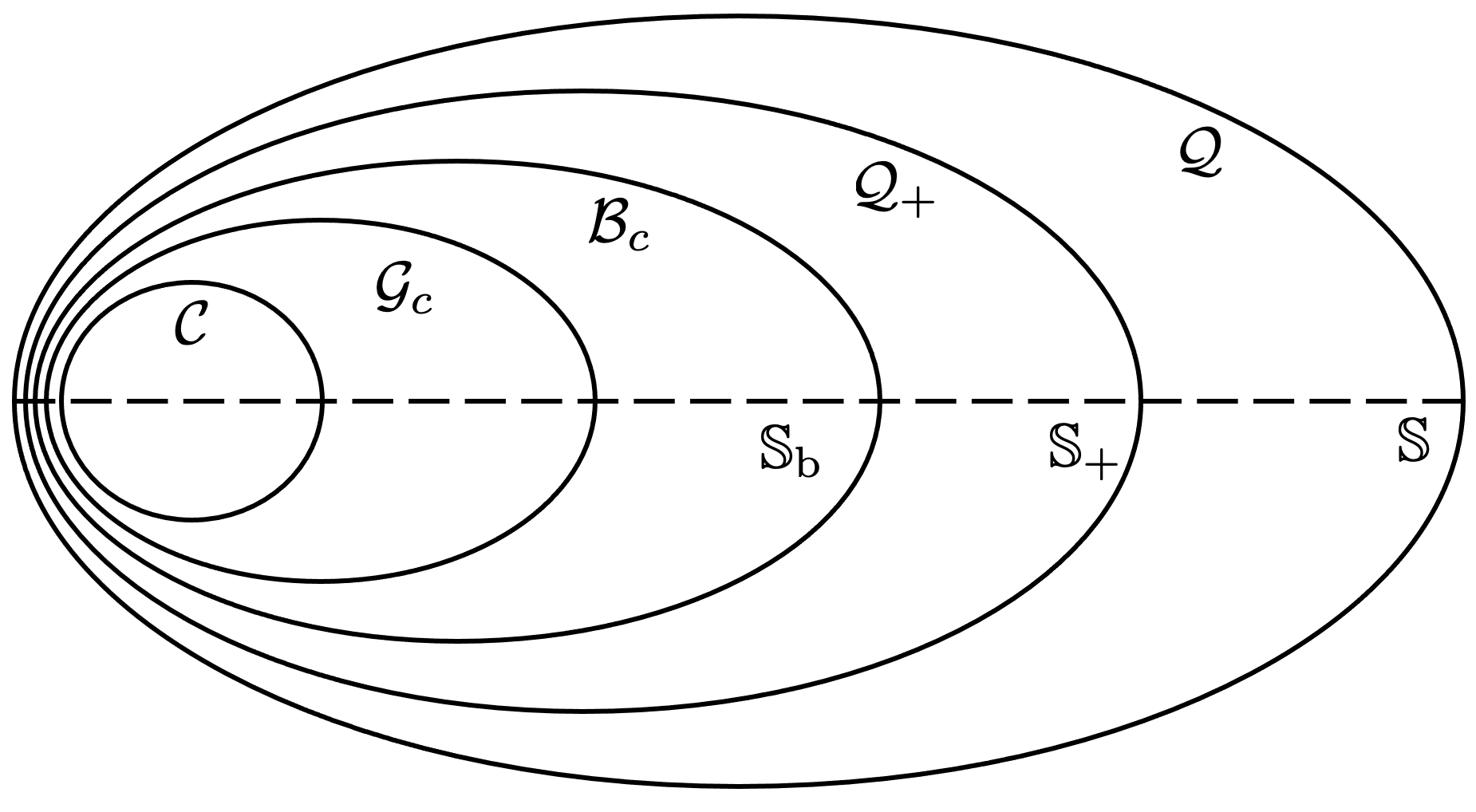}
	\caption{Pictorial representation of the various sets considered here. The full set of quantum states is denoted as $\mathcal{Q}$, while the set of Wigner-positive states is denoted as $\mathcal{Q}_+$. Then, $\mathcal{B}_c$ stands for the convex hull of the set $\mathcal{B}$ of beam-splitter states, while $\mathcal{G}_c$ stands for the convex hull of the set $\mathcal{G}$ of Gaussian states. Further,  $\mathcal{C}$ stands for the set of classical states. Within all these sets, we distinguish the states that are phase invariant, which are characterized by a probability vector $\mathbf{p}\in\mathbb{S}$. For states in $\mathcal{Q}_+$, the vector $\mathbf{p}\in\mathbb{S}_+$, while for states in $\mathcal{B}_c$, the vector $\mathbf{p}\in\mathbb{S}_{\mathrm{b}}$. To be rigorous, we note that it is unknown whether the phase-invariant restriction of $\mathcal{B}_c$ might also contain some states such that $\mathbf{p}\notin\mathbb{S}_{\mathrm{b}}$. We have rigorously proven this is not the case for states up to two photons only (see below). Note also that the areas of all the above sets should not be understood quantitatively as they are arbitrary and only meant here to illustrate the chain of inclusion. }
	\label{fig:quantum_sets}
\end{figure}

It can be shown with a simple argument that the set of Gaussian states $\mathcal{G}$ is a subset of $\mathcal{B}$.
Indeed, it is well known that the product of two identical copies of a Gaussian state $\hat{\gamma}$ is invariant under the action of a beam splitter (assuming the coherence vector vanishes \cite{Weedbrook2012}). We have the identity $\hat{U}_{\eta}\left(\hat{\gamma}\otimes\hat{\gamma}\right)\hat{U}_{\eta}^\dagger = \hat{\gamma}\otimes\hat{\gamma}$, where $\hat{\gamma}$ is any single-mode Gaussian state and $\hat{U}_\eta$ is the unitary of a beam splitter with transmittance $\eta$. One can then easily reconstruct the set of Gaussian states with the above setup, so it follows that $\mathcal{G}\subset\mathcal{B}$. Note that it is easy to build beam-splitter states as in Fig. \ref{fig:bbs_schema} that are not Gaussian states; hence this is a strict inclusion relation. The analog relation also applies to the respective convex hull of these sets, namely, $\mathcal{G}_c\subset\mathcal{B}_c$. Unfortunately, the set $\mathcal{B}_c$ does not coincide with $\mathcal{Q}_+$ as we will see that there exist Wigner-positive states that do not belong to $\mathcal{B}_c$ (see, e.g., the dark blue region in Fig.~\ref{fig:wig_pos_2}). In summary, we have the following chain of strict inclusion relations:
\begin{equation}
	\mathcal{Q}\supset\mathcal{Q}_+\supset\mathcal{B}_c\supset\mathcal{G}_c\supset\mathcal{C} \,
\end{equation}
as pictured in Fig. \ref{fig:quantum_sets}.

\subsection*{Phase-invariant states in $\mathcal{Q}_+$}

As it appears, the set $\mathcal{Q}_+$ of Wigner-positive states remains hard to encompass and characterize efficiently. Therefore, in order to make a concrete step towards the proof of conjecture \eqref{eq:wig_conj}, we restrict our attention in this paper to a class of quantum states known as phase-invariant states. Phase-invariant states have a Wigner function that is invariant under rotation, so they are fully characterized by their radial Wigner function. Such states have the advantage of being easily characterized in state space as they can be written as mixtures of Fock states, which are eigenstates of the harmonic oscillator.
The wave function and the Wigner function of the $n^{\mathrm{th}}$ Fock state (starting at $n=0$ for vacuum) are the following:
\begin{align} 
	\psi_n(x) &=   \pi^{-\frac{1}{4}}2^{-\frac{n}{2}}\left(n!\right)^{-\frac{1}{2}}H_n(x)\exp\left(-\frac{x^2}{2}\right),
	 \label{eq:wave_function_fock}
	 \\
	W_n(x,p) &=    \frac{1}{\pi}(-1)^n L_n\left(2x^2+2p^2\right)\exp\left(-x^2-p^2\right),
	\label{eq:wigner_function_fock}
\end{align}
where $H_n$ and $L_n$ are respectively the $n^{\mathrm{th}}$ Hermite and Laguerre polynomials. A phase-invariant state is thus expressed as the mixture
\begin{equation}
\hat{\rho} = \sum_{k=0}^{\infty} p_k \ket{k}\bra{k}
\end{equation}
with $\ket{k}$ denoting the $k^{\mathrm{th}}$ Fock state, so that it is fully described by the probability vector $\mathbf{p}\in\mathbb{R}^{\mathbb{N}}$, with components $p_k$. In order to be an acceptable probability distribution, $\mathbf{p}$ must satisfy the physicality conditions
\begin{equation}
	p_k\geq 0\quad\forall k,
	\qquad
	\sum\limits_{k=0}^{\infty}p_k = 1.
	\label{eq:physicality}
\end{equation}
We call $\mathbb{S}$ the restriction of $\mathbb{R}^{\mathbb{N}}$ satisfying the physicality conditions \eqref{eq:physicality}.
Any vector $\mathbf{p}$ that belongs to $\mathbb{S}$ corresponds to a unique phase-invariant state in $\mathcal{Q}$.

Now, we turn to the phase-invariant states in $\mathcal{Q}_+$. In order to check that the phase-invariant state that is characterized by a vector $\mathbf{p}\in\mathbb{S}$ is Wigner-positive, we need to verify that the corresponding mixture of Fock states has a positive Wigner function everywhere in phase space. This is done by using Eq. \eqref{eq:wigner_function_fock}, so that the Wigner-positivity condition on $\mathbf{p}$ reads as
\begin{equation}
	\sum\limits_{k=0}^{\infty}
	p_k \, (-1)^k \, L_k(t)
	\geq 0\qquad\forall t\geq 0,
	\label{eq:wigner_positivity}
\end{equation}
where we define $t=2x^2+2p^2$.
Let us also define the usual radial parameter $r=\sqrt{x^2+p^2}$, so that each value of $t$ corresponds to a specific value of $r$ through the relation $t=2r^2$.
When condition \eqref{eq:wigner_positivity} is fulfilled for some $t$, the Wigner function is non-negative at $r=\sqrt{t/2}$.
We call $\mathbb{S}_+$ the restriction of $\mathbb{S}$ satisfying the Wigner-positivity conditions \eqref{eq:wigner_positivity}, so that any vector $\mathbf{p}$ in $\mathbb{S}_+$ is associated with a unique phase-invariant Wigner-positive state in $\mathcal{Q}_+$.

The characterization of $\mathbb{S}_+$ can be operated as follows. Each value of $t$ in Eq. \eqref{eq:wigner_positivity} gives the equation of a hyperplane dividing $\mathbb{S}$ in two halves [$\mathbf{p}$ must be located on one side of the hyperplane to guarantee that $W(r)\ge 0$ for the corresponding $r$]. Two hyperplanes associated respectively to $t$ and $t+\mathrm{d}t$ intersect in a (lower-dimensional) hyperplane which is at the boundary of the convex set $\mathbb{S}_+$.
When $t$ goes from $0$ to $\infty$, the collection of all these intersections forms a locus of points which determines the curved boundary of $\mathbb{S}_+$.
Mathematically, the condition that a point $\mathbf{p}\in\mathbb{S}$ belongs to the curved boundary of $\mathbb{S}_+$ is equivalent to the following condition:
\begin{equation}
	\exists t \quad\text{such that}\quad
	\begin{cases}
		\sum\limits_{k=0}^{\infty} p_k \, (-1)^k \, L_k(t) = 0
		\\[1.2em]
		\sum\limits_{k=0}^{\infty} p_k \, (-1)^k \, \dfrac{\mathrm{d}}{\mathrm{d}t}L_k(t) = 0
	\end{cases}
	\label{eq:boundary_nonflat}
\end{equation}
Note that since $\mathbb{S}_+$ is convex, all the points in its curved boundary are extremal points.
However, other isolated extremal points may exist, as illustrated in Fig. \ref{fig:convex_set_extremal_points}.

\begin{figure}
	\includegraphics[width=6cm]{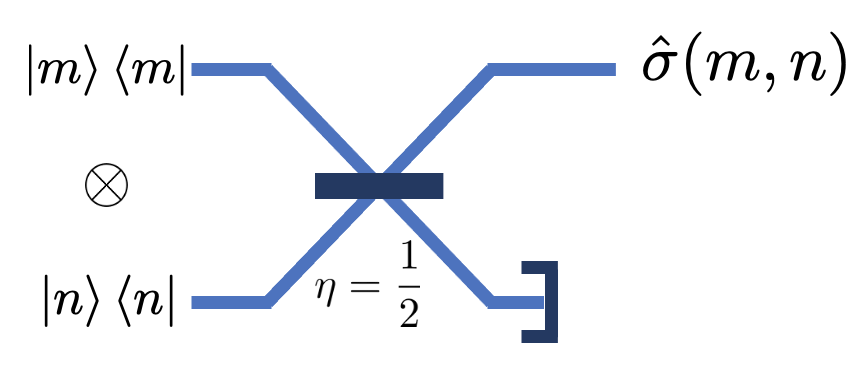}
	\caption{Beam-splitter state ${\hat\sigma}(m,n)$ obtained at the output of a balanced beam splitter (of transmittance $\eta=\frac{1}{2}$) that is fed with Fock states of $m$ and $n$ photons. The states ${\hat\sigma}(m,n)$ are Wigner-positive phase-invariant states; hence they belong to the set $\mathbb{S}_+$.}
	\label{fig:sigma-states}
\end{figure}

\subsection*{Phase-invariant beam-splitter states in $\mathcal{B}$}

The above considerations reflect the fact that characterizing the set of phase-invariant Wigner-positive states (associated with $\mathbf{p}\in\mathbb{S}_+$) remains complex. For this reason, we consider a subset of states that are built by using a balanced beam splitter, following the same idea as for the construction of set $\mathcal{B}$ but injecting phase-invariant Fock states at the input. As pictured in Fig. \ref{fig:sigma-states}, we define the beam-splitter state $\hat{\sigma}(m,n)$ as the reduced output state of a balanced beam splitter fed by $m$ and $n$ photons at its two inputs, that is,
\begin{equation}
	\hat{\sigma}(m,n) = 
	\mathrm{Tr}_{2}
	\left[
	\hat{U}_{\frac{1}{2}}
	\left(
	\ket{m}\bra{m}
	\otimes
	\ket{n}\bra{n}
	\right)
	\hat{U}_{\frac{1}{2}}^\dagger
	\right].
	\label{eq:def_sigma_bs}
\end{equation}
Thus, any state $\hat{\sigma}(m,n)$ is Wigner positive and phase invariant. It is a mixture of Fock states with mixture coefficients given in Appendix \ref{apd:bbs_wigner_positive}. We denote as $\mathbb{S}_{\mathrm{b}}$ the set of probability vectors $\mathbf{p}$ corresponding to all mixtures of states $\hat{\sigma}(m,n)$. It is clear that  $\mathbb{S}_{\mathrm{b}} \subset \mathbb{S}_{+}\subset \mathbb{S}$, as depicted in Fig. \ref{fig:quantum_sets} and discussed below.

Interestingly, the Wigner function associated with any state $\hat{\sigma}(m,n)$ happens to have a minimum value that reaches precisely zero [except for $\hat{\sigma}(0,0)$, which is simply the vacuum state].  In fact, it is shown in Appendix \ref{apd:bbs_wigner_positive} that whenever $m\neq n$, the Wigner function of $\hat{\sigma}(m,n)$ always cancels at the origin in phase space. This suggests that the states $\hat{\sigma}(m,n)$ are the extremal states of the set of phase-invariant Wigner-positive states (those associated with $\mathbb{S}_{+}$). However, as we will show in the following example, the situation is more tricky as this set also admits other extremal states that are not of the form $\hat{\sigma}(m,n)$. Hence, we will see that $\mathbb{S}_{\mathrm{b}} \subset \mathbb{S}_{+}$ is a strict inclusion and there exist phase-invariant Wigner-positive states that cannot be written as mixtures of beam-splitter states $\hat{\sigma}(m,n)$.

\subsection*{Example : restriction to two photons}

\begin{figure}
	\includegraphics[width=7cm]{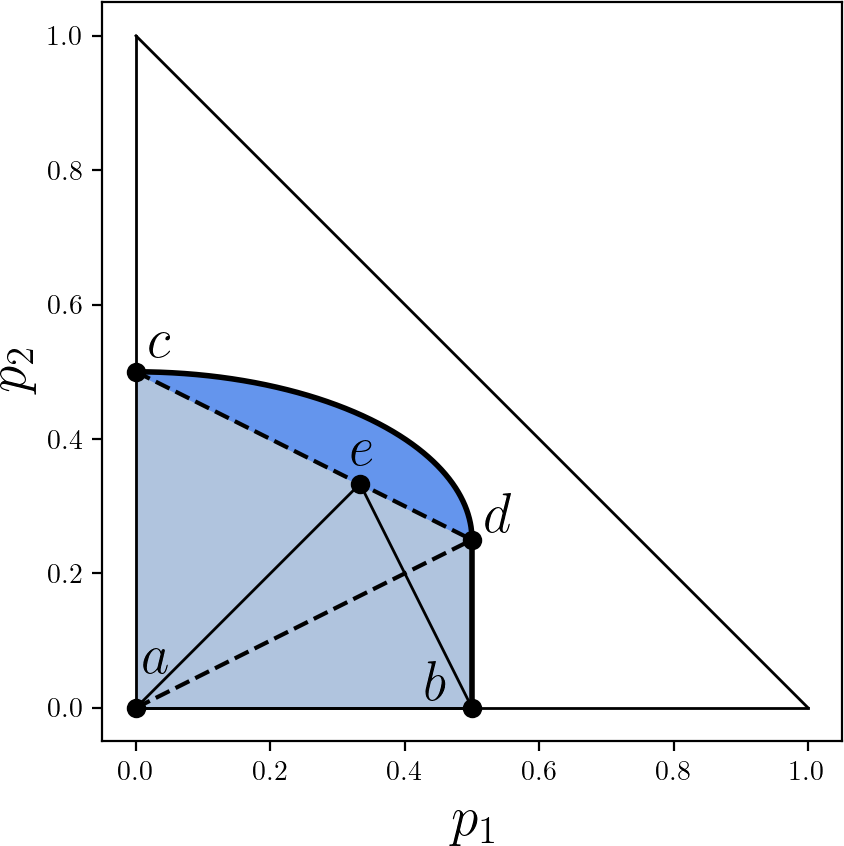}
	\caption{Two-dimensional representation of $\mathbb{S}^2$ (large white triangle) and $\mathbb{S}^{2}_{+}$ (blue zone, including dark and light blue).  The points $a$, $b$, $c$, and $d$ correspond to the beam-splitter states whose convex closure yields $\mathbb{S}^{2}_{\mathrm{b}}$, represented as the light blue zone. The dark blue zone stands for the subset of phase-invariant Wigner-positive states that cannot be expressed as mixtures of beam-splitter states. As discussed in Sec.~\ref{sec:results}, the triangle $a$-$b$-$e$ encompasses the set of passive states while the triangle $a$-$b$-$d$ encompasses the states whose Wigner function coincides with the Husimi $Q$ function of a state.}
	\label{fig:wig_pos_2}
\end{figure}

\begin{figure}
	\includegraphics[width=7.5cm]{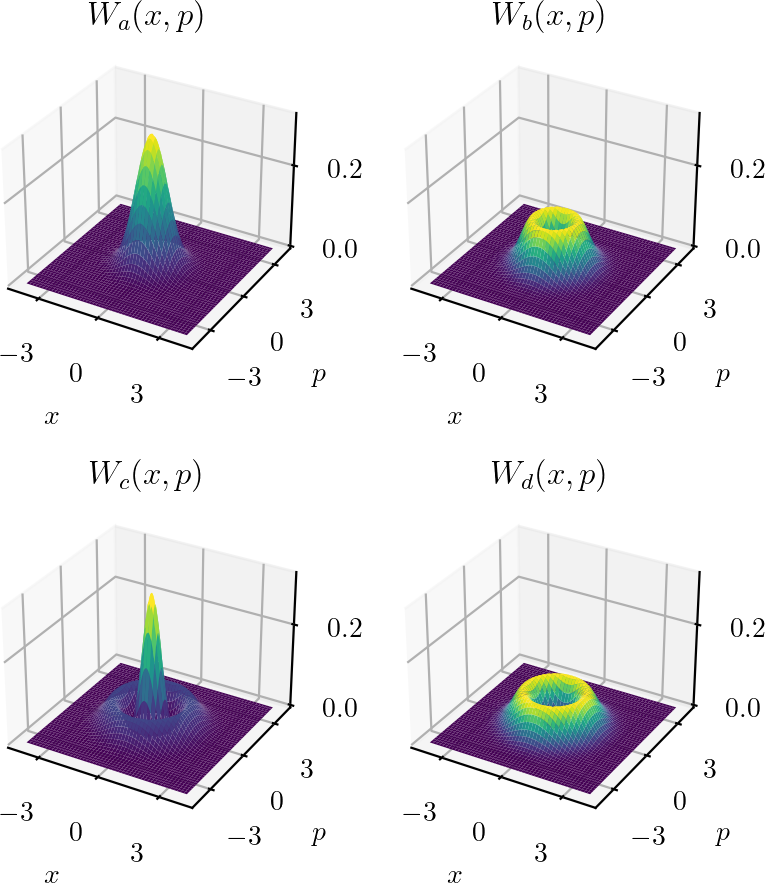}
	\caption{Wigner functions of the four beam-splitter states $\hat{\sigma}_a$, $\hat{\sigma}_b$, $\hat{\sigma}_c$, and $\hat{\sigma}_d$, denoted respectively as $W_a(x,p)$, $W_b(x,p)$, $W_c(x,p)$, and $W_d(x,p)$. These four states are Wigner-positive phase-invariant states, but, in addition, their Wigner functions touch precisely zero (except for the vacuum state $\hat{\sigma}_a$) as is more evident from Fig. \ref{fig:wigner_sigma_states_radial}. This fact reflects that these are extremal states of the set of Wigner-positive phase-invariant  states (associated with $\mathbb{S}^2_+$).  }
	\label{fig:wigner_sigma_states}
\end{figure}

\begin{figure}
	\includegraphics[width=8cm]{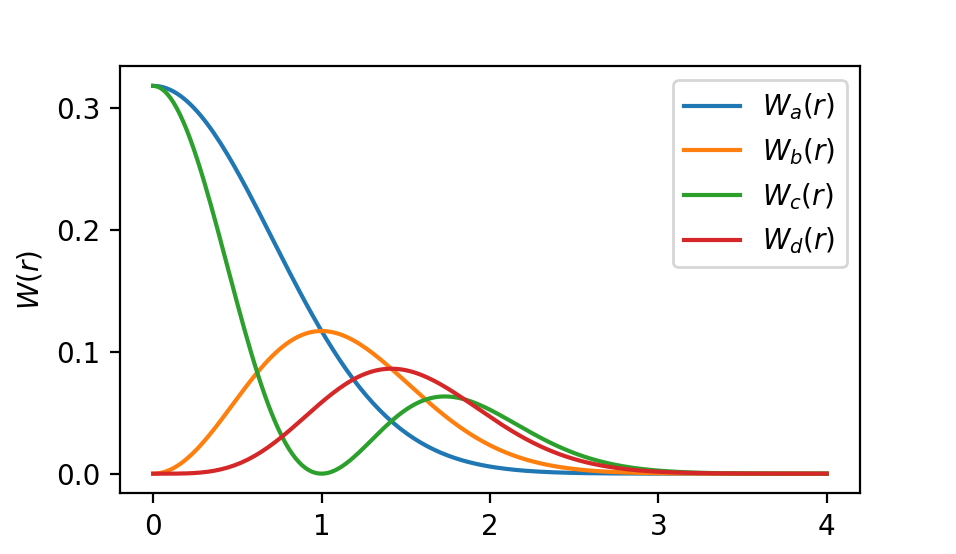}
	\caption{Radial Wigner functions of the four phase-invariant beam-splitter states $\hat{\sigma}_a$, $\hat{\sigma}_b$, $\hat{\sigma}_c$, and $\hat{\sigma}_d$, denoted respectively as $W_a(r)$, $W_b(r)$, $W_c(r)$, and $W_d(r)$. As advertised, the minimum value of these Wigner functions touches zero  [except for the vacuum state $\hat{\sigma}_a$, for which $W_a(r)\to 0$ as $r\to\infty$]. }
	\label{fig:wigner_sigma_states_radial}
\end{figure}

Let us denote by $\mathbb{S}^n$ and $\mathbb{S}^{n}_{+}$ the restriction of respectively $\mathbb{S}$ and $\mathbb{S}_{+}$ that have components $p_k = 0$ for $k>n$. 
As an example, let us consider the set $\mathbb{S}^2$, which corresponds to mixtures of Fock states up to $n=2$, that is,
\begin{equation}
	\hat{\rho} = 
	(1-p_1-p_2)\ket{0}\bra{0}+
	p_1\ket{1}\bra{1}+
	p_2\ket{2}\bra{2}
	\label{eq:rho_mixt_2fock}
\end{equation}
with $p_1,p_2\ge 0$ and $p_1+p_2\le 1$.
We are interested in the Wigner-positive subset of $\mathbb{S}^2$, namely, $\mathbb{S}^2_+$.
Restricting ourselves to $n=2$ makes it possible to represent $\mathbb{S}^2_+$ in a two-dimensional plane with coordinates $p_1$ and $p_2$ (see Fig. \ref{fig:wig_pos_2}).
The mathematical description of $\mathbb{S}^2_+$ was also given in \cite{Brocker1995}, but we analyze it here from a physical perspective, through the prism of quantum optics.
Since the beam splitter conserves the total photon number, we know that only the states $\hat{\sigma}(m,n)$ such that $m+n\leq 2$ belong to $\mathbb{S}^2_+$.
These states are expressed as
\begin{equation}
	\begin{split}
		\hat{\sigma}_a &\equiv \hat{\sigma}(0,0) = \hspace{0.9em}\ket{0}\bra{0}
		\\
		\hat{\sigma}_b &\equiv  \hat{\sigma}(1,0) =
		\frac{1}{2}\ket{0}\bra{0}+\frac{1}{2}\ket{1}\bra{1}
		\\
		\hat{\sigma}_c &\equiv  \hat{\sigma}(1,1) =
		\frac{1}{2}\ket{0}\bra{0}+\frac{1}{2}\ket{2}\bra{2}
		\\
		\hat{\sigma}_d &\equiv  \hat{\sigma}(2,0) =
		\frac{1}{4}\ket{0}\bra{0}+\frac{1}{2}\ket{1}\bra{1}+\frac{1}{4}\ket{2}\bra{2}
	\end{split}
\end{equation}
and their corresponding Wigner functions are displayed in Figs. \ref{fig:wigner_sigma_states} and \ref{fig:wigner_sigma_states_radial}. We observe that the minimum value of the Wigner functions always reaches zero (except for the vacuum state ${\hat \sigma}_a$), which reflects that these are extremal states of the set of Wigner-positive phase-invariant states (associated with $\mathbb{S}^2_+$).

This is confirmed in Fig. \ref{fig:wig_pos_2}, where the four beam-splitter states are represented by points $a$, $b$, $c$, and $d$: they are indeed extremal points of the convex set $\mathbb{S}^2_{+}$, which appears as the blue zone (including light and dark blue).  However, as we will see, they are not the only extremal points of  $\mathbb{S}^2_{+}$.
The complete characterization of $\mathbb{S}^2_+$ can be done by using the Wigner-positivity conditions \eqref{eq:wigner_positivity} and \eqref{eq:boundary_nonflat}. The derivation is done in Appendix \ref{apd:mixture_2phot} and leads to the following conditions on $p_1$ and $p_2$:
\begin{equation}
	\begin{cases}
		p_1\leq \dfrac{1}{2}
		\\[1em]
		p_2\leq \dfrac{1}{4}+\dfrac{1}{4}\sqrt{1-4p_1^2}
	\end{cases}
	\label{eq:domain_S2plus}
\end{equation}
Any state in the form \eqref{eq:rho_mixt_2fock} is Wigner positive if and only if its components $p_1$ and $p_2$ satisfy conditions \eqref{eq:domain_S2plus}.

\begin{figure}
	\includegraphics[width=8cm]{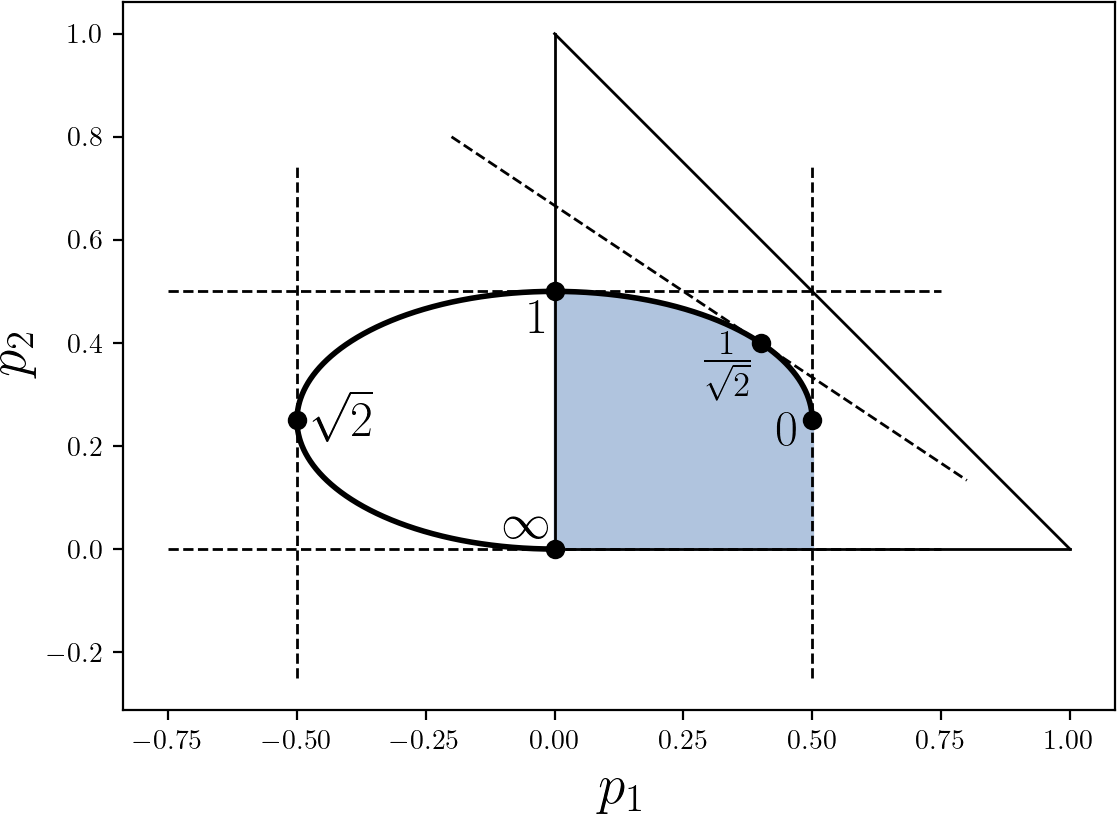}
	\caption{Expressing the positivity of the radial Wigner function $W(r)$ for increasing values of $r$ corresponds to a continuum of straight lines, which are all tangents of ellipse \eqref{eq:ellipse}. As an illustration, we plot as dashed lines the tangents associated with $W(r)=0$ for $r=0$, $r=1/\sqrt{2}$, $r=1$, $r=\sqrt{2}$, and $r\to \infty$. For instance, expressing $W(0)\ge 0$ implies $p_1\le 1/2$, while expressing $W(1)\ge 0$ implies $p_2\le 1/2$. For $r>1$, the positivity condition becomes redundant, and, at the limit $r\to\infty$, it gives $p_2\ge 0$, which is equivalent to the physicality condition.}
	\label{fig:rotating-tangent}
\end{figure}

Several observations can be made from Fig. \ref{fig:wig_pos_2}. First, state ${\hat \sigma}_a$, which coincides with the vacuum state, is a trivial extremal state of $\mathbb{S}^2_+$ even if its Wigner function does not reach zero. As already mentioned, ${\hat \sigma}_b$, ${\hat \sigma}_c$, and ${\hat \sigma}_d$ are other extremal states of $\mathbb{S}^2_+$, as witnessed by the fact that their Wigner function vanishes at some location in phase space. The convex set $\mathbb{S}^2_+$ has three facets.
Two of them correspond to the physicality conditions \eqref{eq:physicality}, i.e., $p_1\geq 0$ and $p_2\geq 0$.
The third one corresponds to condition \eqref{eq:wigner_positivity} where we have set $t=0$, which gives us $p_0+p_2\geq 1/2$ or equivalently $p_1\leq 1/2$. Note that the points in these facets belong to the boundary of $\mathbb{S}^2_+$ but are not extremal. 
This can be easily understood for the third facet corresponding in Fig. \ref{fig:wig_pos_2}  to the segment connecting ${\hat \sigma}_b$ to ${\hat \sigma}_d$, which both admit a zero of their Wigner function at the same location (i.e., the origin). Note also that, in general, the set $\mathbb{S_+}$ always has a facet corresponding to 
\begin{equation}
	\sum\limits_{k\ \mathrm{even}} p_k= \dfrac{1}{2},
	\label{eq:boundary_flat}
\end{equation}
which expresses the positivity of the Wigner function at $r=0$ (recall that $t = 2r^2$). 
As pictured in Fig. \ref{fig:rotating-tangent}, expressing the positivity of the radial Wigner function for increasing values of $r$ yields a continuum of straight lines, whose locus of intersecting points forms an ellipse centered in $(0,1/4)$, namely,
\begin{equation}
\left(\frac{p_1}{1/2}\right)^2 + \left(\frac{p_2-1/4}{1/4}\right)^2 = 1 \,  .
\label{eq:ellipse}
\end{equation}
 The resulting constraints on $p_1$ and $p_2$ for all $r$'s are summarized by  Eq. \eqref{eq:domain_S2plus}.

Overall, Fig. \ref{fig:wig_pos_2} shows that the subspace $\mathbb{S}^{2}_{\mathrm{b}}$, which is spanned by the extremal states ${\hat \sigma}_a$, ${\hat \sigma}_b$, ${\hat \sigma}_c$, and ${\hat \sigma}_d$, covers a large region of $\mathbb{S}^{2}_{+}$ (indicated in light blue) so any point in this region can thus be generated by a convex mixture of them. However, $\mathbb{S}^{2}_{+}$ also includes a small region (indicated in dark blue) that is located under the ellipse defined by Eq. \eqref{eq:ellipse}  and above the straight line $c$-$d$. This region is thus outside the polytope $\mathbb{S}^{2}_{\mathrm{b}}$ generated by the $\hat{\sigma}$ states, which confirms that $\mathbb{S}^{2}_{+}$ also admits a continuum of extremal points along this ellipse.

Note finally that it is not a trivial observation to see that $\mathbb{S}^2_{\mathrm{b}}$ coincides with the two-photon phase-invariant restriction of $\mathcal{B}_c$ (i.e., the phase-invariant states with up to two photons within the convex hull of beam-splitter states of $\mathcal{B}$). Indeed, $\mathbb{S}^2_\mathrm{b}$ is defined as the convex hull of beam-splitter states built from (phase-invariant) Fock states  in Fig.~\ref{fig:sigma-states} with up to two photons, that is, the convex hull of $\lbrace \hat{\sigma}_a,\hat{\sigma}_b,\hat{\sigma}_c,\hat{\sigma}_d\rbrace$.
Since it is possible to create beam-splitter states in the setup of Fig.~\ref{fig:bbs_schema} that are phase-invariant starting from two input states that are not phase invariant (e.g., two squeezed states with orthogonal squeezing produce a thermal state), it might  \textit{a~priori} be possible to build states within the two-photon phase-invariant restriction of $\mathcal{B}_c$ that do not belong to $\mathbb{S}^2_\mathrm{b}$. However, a simple argument convinces us otherwise. First, notice that we may restrict to pure input states without loss of generality. Since the output is a mixture with up to two photons, we must consider input states that are either in the form
\begin{equation}
	\ket{\psi}=\ket{0} \otimes \left(a_0\ket{0}+a_1\ket{1}+a_2\ket{2}\right),
	\label{eq_first-case}
\end{equation}
or
\begin{equation}	
	\ket{\psi}=\left(b_0\ket{0}+b_1\ket{1}\right)\otimes\left(c_0\ket{0}+c_1\ket{1}\right).
	\label{eq_second-case}
\end{equation}
In case \eqref{eq_first-case}, the first input is the vacuum, which is phase invariant, so that the output state is phase invariant only if the second input state is also phase invariant. This is easy to understand given that the output Wigner function is a (scaled) convolution of the two input Wigner functions. In case \eqref{eq_second-case}, a straightforward calculation shows us that the output state is phase invariant only if at least one of the coefficients $b_0$, $b_1$, $c_0$, or $c_1$ vanishes.
This implies that one of the two input states must be phase invariant, which in turns implies that the other input must be phase-invariant too in order to ensure the phase invariance of the output. As a result, the  two-photon phase-invariant restriction of $\mathcal{B}_c$ coincides with the set $\mathbb{S}^2_\mathrm{b}$ (it is unknown, however, whether this remains true for more than two photons, that is, whether the phase-invariant restriction of $\mathcal{B}_c$ corresponds to the set $\mathbb{S}_\mathrm{b}$ in general). Since we have found phase-invariant Wigner-positive states outside $\mathbb{S}^{2}_{\mathrm{b}}$, this confirms that $\mathcal{B}_c$ is strictly included in $\mathcal{Q}_+$, as advertised earlier (see Fig. \ref{fig:quantum_sets}).

\section{Conjectured lower bound}
\label{sec:results}

The conjectured lower bound on the Wigner entropy reads
\begin{equation}
	h\left(W_{\! \hat{\rho}}\right)\geq\ln\pi+1 \qquad  \forall \hat{\rho}\in \mathcal{Q}_+ 
	\label{eq:to-be-proven}
\end{equation}
Note that an extended version for the Wigner-R\' enyi entropy is also discussed in Appendix \ref{sect-wigner-renyi}. We wish to prove Eq. \eqref{eq:to-be-proven} for all Wigner-positive states in $ \mathcal{Q}_+$ but it appeared in Sec. \ref{sec:wig_pos} that this set is hard to characterize. In this section, we will expose the central result of our paper, namely, a proof of this conjecture over a subset of phase-invariant Wigner-positive states with thermodynamical relevance that are called \textit{passive states}.
As a side result, we will exhibit an unexpectedly simple relation between the extremal passive states and the beam-splitter states $\hat{\sigma}(m,n)$, which guides us to test the conjecture over the much larger set $\mathbb{S}_{\mathrm{b}}$ of phase-invariant Wigner-positive states.

Before doing so, let us discuss the implication of the conjecture in the restricted subspace of phase-invariant Wigner-positive states associated with $\mathbb{S}^{2}_{+}$. First, as a consequence of Eq. \eqref{eq:fundamental}, we know that the Wigner functions of ${\hat \sigma}_a$, ${\hat \sigma}_b$, and ${\hat \sigma}_d$ coincide respectively with the Husimi $Q$ functions of $\ket{0}$, $\ket{1}$, and $\ket{2}$. Hence, the (proven) Wehrl conjecture applied to $\ket{0}$, $\ket{1}$, and $\ket{2}$ implies that the Wigner entropy of  ${\hat \sigma}_a$, ${\hat \sigma}_b$, and ${\hat \sigma}_d$  is indeed lower bounded by $\ln\pi+1$. Further, this naturally extends to the subspace spanned by ${\hat \sigma}_a$, ${\hat \sigma}_b$, and ${\hat \sigma}_d$, corresponding to the triangle $a$-$b$-$d$ in Fig. \ref{fig:wig_pos_2}. Thus, the states that are located in the blue region but do not belong to this triangle are Wigner-positive states whose Wigner function cannot be expressed as a physical $Q$ function. This underlies the fact that conjecture \eqref{eq:wig_conj} is stronger than the Wehrl conjecture. In particular, let us prove that the Wigner function of state ${\hat \sigma}_c$ cannot be written as the $Q$ function of a physical state. Reasoning by contradiction, assume there exists an input state $\hat{\rho}$ in the setup of Fig.~\ref{fig:bbs_vacuum} such that the resulting output state is ${\hat \sigma}_c$. First, since the transformation on $\hat{\rho}$ is a (scaled) convolution with a (Gaussian) rotation-invariant function, the Wigner function of $\hat{\rho}$ must necessarily be rotation invariant in order to get the rotation-invariant Wigner function associated with ${\hat \sigma}_c$. Thus, $\hat{\rho}$ must be phase invariant, that is, a mixture of Fock states. Second, since ${\hat \sigma}_c$ does not contain more than two photons, it is clear that $\hat{\rho}$ can only be a mixture of  $\ket{0}$, $\ket{1}$, and $\ket{2}$. However, the output state corresponding to any such mixture precisely belongs to the triangle $a$-$b$-$d$, which does not contain $c$. Hence, there is no state $\hat{\rho}$.

\subsection*{Passive states}

Passive states are defined in quantum thermodynamics as the states from which no work can be extracted through unitary operations \cite{Pusz1978}. 
If $\hat{\rho}_p$ is the density operator of a passive state, then the following relation holds true for any unitary operator $\hat{U}$:
\begin{equation}
	\mathrm{Tr}
	\left[
	\hat{\rho}_p \hat{H}
	\right]
	\leq
	\mathrm{Tr}
	\left[
	\hat{U}\hat{\rho}_p \hat{U}^\dagger \hat{H}
	\right],
\end{equation}
where $\hat{H}$ is the Hamiltonian of the system. Passive states are useless in the sense that it is not possible to decrease their energy by applying a unitary (since a unitary conserves the entropy, any work extraction should come with a decrease of internal energy). 
It can be shown that passive states are decreasing mixtures of energy eigenstates, in the sense that if the eigenstates are labeled with increasing energy, the associate probabilities must be decreasing \cite{Lenard1978}. In the present paper, we are considering eigenstates of the harmonic oscillator, which are the Fock states.
A passive state is then written as 
\begin{equation}
\hat{\rho}_p = \sum_{k=0}^{\infty} p_k \ket{k}\bra{k}   \qquad \mathrm{with~} p_k\geq p_{k+1} .
\label{eq:def_passive_states_in_Fock}
\end{equation}

Among the set of passive states, \textit{extremal} passive states are defined as equiprobable mixtures of the low-energy eigenstates up to some threshold.
We refer to the $n^{\mathrm{th}}$ extremal passive state as $\hat{\varepsilon}_n$ and to its Wigner function as $E_n$.
They are defined as follows:
\begin{equation}
	\begin{split}
		\hat{\varepsilon}_n &= \dfrac{1}{n+1}\sum\limits_{k=0}^{n}\ket{k}\bra{k}
		\\[1em]
		E_n(x,p)&=\dfrac{1}{n+1}\sum\limits_{k=0}^{n}
		W_k(x,p)
		\label{eq:def_extremal_states}
	\end{split}
\end{equation}
The states $\hat{\varepsilon}_n$ are called extremal \footnote{Note that the extremal passive states $\hat{\varepsilon}_n$ are very different from the extremal Wigner-positive states, such as the states $\hat{\sigma}(m,n)$: the extremality of $\hat{\varepsilon}_n$ pertains to a set of states that are defined in state space, while the extremality of $\hat{\sigma}(m,n)$ pertains to a distinct set of states that are defined in phase space.}  in the sense that any passive state $\hat{\rho}_p $ can be expressed as a unique convex mixture of extremal passive states, namely,
\begin{equation}
	\hat{\rho}_p = \sum\limits_{k=0}^{\infty}  e_k \, \hat{\varepsilon}_k  \,  ,
	\label{eq:passive-in-terms-of-extremal-passive}
\end{equation}
where $p_k$ and $e_k$ are probabilities that are linked through the relation $e_k = (k+1)\left(p_k-p_{k+1}\right)$.

In the special case of phase-invariant states within the restricted space with up to two photons, the set of passive states corresponds to the triangle $a$-$b$-$e$ in Fig. \ref{fig:wig_pos_2}, which belongs to $\mathbb{S}^2_+$ as expected. Of course, $a$, $b$, and $e$ correspond respectively to the extremal passive states $\hat{\varepsilon}_0$, $\hat{\varepsilon}_1$, and $\hat{\varepsilon}_2$.


\subsection*{Proof of the conjecture for passive states}


Let us prove the lower bound \eqref{eq:to-be-proven} for the subset of passive states $\hat{\rho}_p$.
First, note that passive states are known to be Wigner positive \cite{Bastiaans1983}, a fact that will become clear from Eq. \eqref{eq:extremal_states_formula}. Thus, their Wigner entropy is well defined. Second, notice that, as a consequence of the concavity of entropy, it is sufficient to prove the conjecture for all extremal passive states $\hat{\varepsilon}_n$.


The main tool that we will use to carry out our proof is a formula that we have derived from an identity involving Laguerre and Hermite polynomials \cite{Szeg1939}, making a nontrivial link between the Wigner functions and wave functions of the first $n$ Fock states. It reads as follows (to the best of our knowledge, it has never appeared as such in the literature):
\begin{equation}
	\sum\limits_{k=0}^{n}W_k(x,p) = 
	\sum\limits_{k=0}^{n}\psi_{k}(x)^2 \, \psi_{n-k}(p)^2,
	\label{eq:extremal_states_formula}
\end{equation}
where $W_k$ and $\psi_k$ are respectively the Wigner function and wave function of the $k^{\text{th}}$ Fock state as defined in Eqs. \eqref{eq:wave_function_fock} and \eqref{eq:wigner_function_fock}. As a by-product, note that Eq. \eqref{eq:extremal_states_formula} immediately implies that all extremal passive states $\hat{\varepsilon}_n$ admit a positive Wigner function; hence the Wigner function of an arbitrary passive state is necessarily positive.
More details on the derivation of Eq. \eqref{eq:extremal_states_formula} can be found in Appendix \ref{apd:formula_extremal_states}.

Let us denote the $x$ and $p$ probability densities of the $n^\text{th}$ Fock state as $\rho_n(x) = \vert\psi_n(x)\vert^2$ and $\rho_n(p) = \vert\psi_n(p)\vert^2$.
Their corresponding Shannon differential entropy is defined as $h\left(\rho_k(x)\right) = -\int\rho_k(x)\ln\rho_k(x)\, \mathrm{d}x$ and $h\left(\rho_k(p)\right) = -\int\rho_k(p)\ln\rho_k(p)\, \mathrm{d}p$. In the following, we refer to these quantities as $h\left(\rho_k\right) \equiv h\left(\rho_k(x)\right)=h\left(\rho_k(p)\right)$.
We are now ready to lower bound the Wigner entropy of the  $n^{\text{th}}$ extremal passive state $\hat{\varepsilon}_n$ by using Eq. \eqref{eq:extremal_states_formula}:
\begin{equation}
	\begin{split}
		h\left(E_n \right)
		&=
		h\left(\dfrac{1}{n+1} \sum\limits_{k=0}^{n}W_k(x,p) \right)
		\\
		&=
		h\left(\dfrac{1}{n+1}\sum\limits_{k=0}^{n}\psi_k(x)^2\psi_{n-k}(p)^2\right)
		\\
		&\geq
		\dfrac{1}{n+1}
		\sum\limits_{k=0}^{n}
		h\big(\rho_k(x)\rho_{n-k}(p)\big)
		\\
		&=
		\dfrac{1}{n+1}
		\sum\limits_{k=0}^{n}
		\big(h\left(\rho_k\right)+h\left(\rho_{n-k}\right)\big)
		\\
		&=\dfrac{2}{n+1}\sum\limits_{k=0}^{n}
		h\left(\rho_k\right)
		\\
		&\geq \ln\pi+1
	\end{split}
	\label{eq:development_extremal_states_wig_conj}
\end{equation}
The first inequality in Eq. \eqref{eq:development_extremal_states_wig_conj} results from the concavity of the entropy. Then, we use the fact that the entropy of a product distribution is the sum of the marginal entropies.  Finally, we apply the entropic uncertainty relation of Białynicki-Birula and Mycielski \cite{Bialynicki1975} on Fock states, namely, $2\, h\left(\rho_k\right)\geq\ln\pi+1$, $\forall k$. We have thus proven the conjecture for all extremal passive states and this proof naturally extends to the whole set of passive states. $\qed$

Let us now prove that a slightly tighter lower bound can be derived for the Wigner entropy of passive states by exploiting Eq. \eqref{eq:passive-in-terms-of-extremal-passive}, namely the fact that these states can be expressed as convex mixtures of extremal passive states $\hat{\varepsilon}_n$ (in place of decreasing mixtures of Fock states). We denote the Wigner function of the passive state $\hat{\rho}_p$ as $W_P(x,p)$ and bound its Wigner entropy as follows:
\begin{equation}
	\begin{split}
		h\left(W_P\right) &= h\left(\sum\limits_{k=0}^{\infty}e_k \, E_k(x,p)\right)
		\\
		&\geq
		\sum\limits_{k=0}^{\infty}
		e_k \, 
		h\big(E_k(x,p)\big)
		\\
		&=
		\sum\limits_{k=0}^{\infty}
		(k+1)\left(p_k-p_{k+1}\right)
		h\big(E_k(x,p)\big)
		\\
		&\geq
		\sum\limits_{k=0}^{\infty}
		(k+1)\left(p_k-p_{k+1}\right)
		\dfrac{2}{k+1}
		\sum\limits_{j=0}^{k}
		h\left(\rho_j\right)
		\\
		&= 2
		\sum\limits_{k=0}^{\infty}
		\sum\limits_{j=0}^{k}
		\left(p_k-p_{k+1}\right)
		h\left(\rho_j\right)
		\\
		&= 2
		\sum\limits_{j=0}^{\infty}
		\sum\limits_{k=j}^{\infty}
		\left(p_k-p_{k+1}\right)
		h\left(\rho_j\right)
		\\
		&= 2
		\sum\limits_{j=0}^{\infty}
		p_j  \, h\left(\rho_j\right).    \qed
	\end{split} 
\label{eq:second_proof}
\end{equation}
The first inequality in \eqref{eq:second_proof} comes from the concavity of entropy over the convex set of extremal states, while the second inequality is obtained from Eq. \eqref{eq:development_extremal_states_wig_conj}. The final expression is a stronger lower bound on the Wigner entropy of any passive state which reads as
\begin{equation}
	h\left(  \sum\limits_{k=0}^{\infty}
	p_k \, W_k   \right)
	\geq 2
	\sum\limits_{k=0}^{\infty}
	p_k \, h\left(\rho_k\right)
	\label{eq:lower_bound_passive_states}
\end{equation}
and is valid as soon as the probabilities $p_k$ are decreasing, that is, $p_k\geq p_{k+1}$.

It is tempting to extrapolate that  the bound \eqref{eq:lower_bound_passive_states} remains valid beyond the set of passive states.
We know indeed that there exist phase-invariant Wigner-positive states that are not passive states (in Fig. \ref{fig:wig_pos_2}, these are the states within the light blue region that do not belong to the triangle $a$-$b$-$e$). As long as the coefficients $p_k$ are such that the corresponding state is Wigner positive, it has a well-defined Wigner entropy and we may expect that the lower bound \eqref{eq:lower_bound_passive_states} applies. Unfortunately, our numerical simulations have shown that relation \eqref{eq:lower_bound_passive_states} does not hold in general for nonpassive (Wigner-positive) states. Of course, we conjecture that  relation \eqref{eq:to-be-proven} does hold for such states and we have not found any counterexample.

\subsection*{Relation between the extremal passive states and the beam-splitter states}

Let us now highlight an interesting relation between extremal passive states  $\hat{\varepsilon}_n$ and the beam-splitter states $\hat{\sigma}(m,n)$ that we defined in Sec. \ref{sec:wig_pos}. To this purpose, we consider a mixed quantum state of two modes (or harmonic oscillators) which we denote as $\hat{\tau}_n$. It is defined as an equal mixture of all two-mode states with a total photon number (or energy) equal to $n$, namely,
\begin{equation}
	\hat{\tau}_n = 
	\dfrac{1}{n+1}
	\sum\limits_{k=0}^{n}
	\ket{k}\bra{k}\otimes\ket{n-k}\bra{n-k}.
	\label{eq:def_tau_2modes}
\end{equation}
This state is maximally mixed over the set of states with total energy $n$, so that it is invariant under any unitary transformation that preserves the total energy.
In particular, it is invariant under the action of a balanced beam splitter, which implies the identity $\hat{U}_{1/2}\ \hat{\tau}_n \  \hat{U}_{1/2}^\dagger = \hat{\tau}_n$.
After partial tracing over the second mode, we obtain
\begin{equation}
	\mathrm{Tr}_2
	\left[
	\hat{\tau}_n
	\right]
	=
	\dfrac{1}{n+1}
	\sum\limits_{k=0}^{n}
	\ket{k}\bra{k} \,  ,
\end{equation}
which is simply the extremal state $\hat{\varepsilon}_n$.
Alternatively, exploiting the invariance under $\hat{U}_{1/2}$ and recalling the definition of the beam-splitter states $\hat{\sigma}(m,n)$, we have
\begin{equation}
	\mathrm{Tr}_2
	\left[
	\hat{\tau}_n
	\right]
	=
	\dfrac{1}{n+1}
	\sum\limits_{k=0}^{n}
	\hat{\sigma}(k,n-k).
\end{equation}
This establishes an interesting link between the extremal passive states and the beam-splitter states, namely,
\begin{equation}
	 \hat{\varepsilon}_n =
	\dfrac{1}{n+1}
	\sum\limits_{k=0}^{n}
	\hat{\sigma}(k,n-k).
\end{equation}
Expressed in terms of Wigner function, this translates as
\begin{equation}
	\sum\limits_{k=0}^{n}
	W_k(x,p)
	=
	\sum\limits_{k=0}^{n}
	S_{(k,n-k)}
	(x,p),
	\label{eq:sum_fock_sum_sigma}
\end{equation}
where $S_{(m,n)}$ denotes the Wigner function of $\hat{\sigma}(m,n)$.

It is instructive to compare Eq.~\eqref{eq:sum_fock_sum_sigma} with Eq.~\eqref{eq:extremal_states_formula}.
Extremal passive states $\hat{\varepsilon}_n$ are defined as mixtures of Fock states [see Eq.~\eqref{eq:def_passive_states_in_Fock}], which possess each a nonpositive Wigner function (except for the vacuum). This is at the heart of the difficulty of proving the conjecture: we cannot give a meaning to the Wigner entropy of a Fock state (except for the vacuum), so the convex decomposition of a state into Fock states cannot be used to bound its Wigner entropy. In this context, both Eqs. \eqref{eq:extremal_states_formula} and \eqref{eq:sum_fock_sum_sigma}  have the crucial interest to provide the decomposition of the Wigner function of an extremal passive state into a sum of positive functions.
However, with Eq.~\eqref{eq:extremal_states_formula}, these positive functions do not correspond to \textit{physical} Wigner functions. Numerical simulations indeed show that in general $\psi_k(x)^2\psi_{n-k}(p)^2$ is not a physically acceptable Wigner function (it is positive but does not correspond to a positive-semidefinite density operator).
On the contrary, Eq. \eqref{eq:sum_fock_sum_sigma} exhibits the decomposition of an extremal passive state into states $\hat{\sigma}(m,n)$, which are Wigner-positive quantum states as we have shown.

\begin{figure}
	\includegraphics[width=8cm]{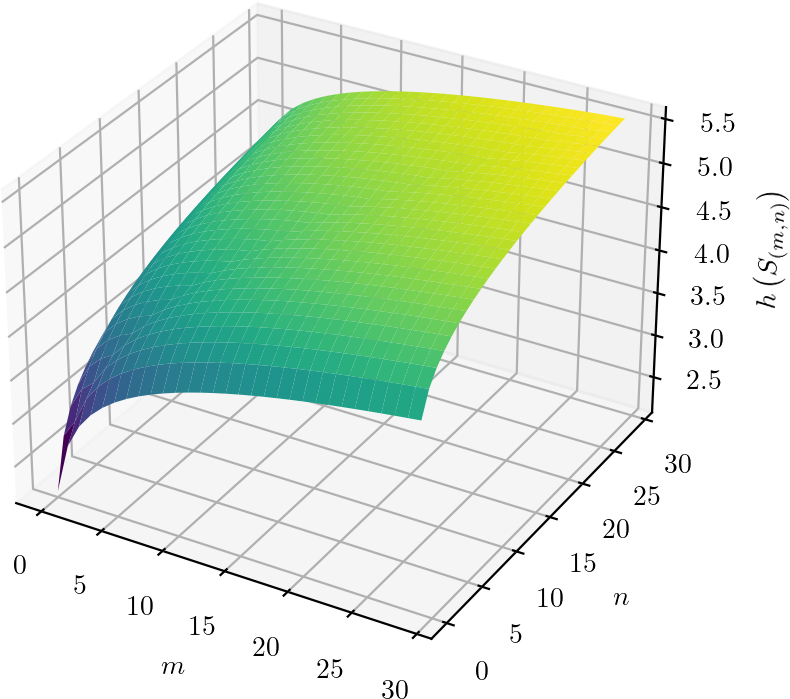}
	\caption{Wigner entropy of the beam-splitter states $\hat{\sigma}(m,n)$ as computed numerically for $m,n=0,1,\dots, \,30$. It appears that the Wigner entropy increases monotonically for increasing values of $m$ and $n$.}
	\label{fig:bbs_entropy}
\end{figure}

The set spanned by the states $\hat{\sigma}(m,n)$ associated with $\mathbb{S}_{\mathrm{b}}$ is obviously bigger than the set of passive states and offers a nice playground for testing our conjecture. Indeed, each state $\hat{\sigma}(m,n)$ is Wigner positive so it has a well-defined Wigner entropy. Figure \ref{fig:bbs_entropy} displays the Wigner entropy of the states $\hat{\sigma}\left(m,n\right)$ as computed numerically up to $m,n=30$. As expected, the minimum Wigner entropy $\ln\pi+1$ is reached for the vacuum state $\sigma\left(0,0\right)=\ket{0}\bra{0}$, so it follows that the conjecture holds for whole set $\mathbb{S}_{\mathrm{b}}$ due the concavity of the entropy. Of course, this is based on numerical evidence since we do not have an analytical proof that $h(S_{(m,n)})\ge \ln\pi+1$.
Further, although the set $\mathbb{S}_{\mathrm{b}}$ is much bigger than the set of passive states, it still does not encompass the whole set of phase-invariant Wigner-positive states  $\mathbb{S}_{+}$, as evidenced by Fig. \ref{fig:wig_pos_2}.

\section{Conclusion}
\label{sec:conclusion}

We have promoted the Wigner entropy of a quantum state as a distinct information-theoretical measure of its uncertainty in phase space. Although it is, by definition, restricted to Wigner-positive states, the fact that such states form a convex set makes it a useful physical quantity. Since it is a concave functional of the state, we naturally turn to its lower bound  over the convex set of Wigner-positive states. We conjecture that this lower bound is $\ln \pi +1$, which is the value taken on by the Wigner entropy of all Gaussian pure states. The latter then play the role of minimum Wigner-uncertainty states. 

This conjecture is consistent with the Hudson theorem, whereby all Wigner-positive pure states must be Gaussian states, thus states reaching the value $\ln \pi +1$. The conjecture also implies a lower bound on the sum of the marginal entropies of $x$ and $p$; hence it results in a tightening of the entropic uncertainty relation due to Bialynicki-Birula and Mycielski that is very natural from the point of view of Shannon information theory (the Wigner entropy accounts for $x$-$p$ correlations since it is the joint entropy of $x$ and $p$). 
Of course, it also implies the Heisenberg uncertainty relation formulated in terms of variances of $x$ and $p$. Furthermore, the conjecture implies (but is stronger than) Wehrl conjecture, notoriously proven by Lieb. It is supported by several elements.  First, we have provided in Sec.  \ref{sec:results} an analytical proof for a subset of phase-invariant Wigner-positive states, namely, the passive states. Second, this was complemented by a semianalytical seminumerical proof for the larger set of phase-invariant states associated with $\mathbb{S}_{\mathrm{b}}$. Third, we also carried out an extensive numerical search for counterexamples in $\mathcal{Q}_+$ but could not find any.

Given that the Wigner entropy is only properly defined for Wigner-positive states, we have also been led to investigate the structure of such states in Sec. \ref{sec:wig_pos}. We have put forward an extensive technique to produce Wigner-positive states using a balanced beam splitter. In particular, we have focused on the beam-splitter states $\hat{\sigma}(m,n)$ and have highlighted their connection with the (smaller) set of passive states and (larger) set of phase-invariant Wigner-positive states. We have also found an unexpectedly simple relation between the states $\hat{\sigma}(m,n)$ and the extremal passive states.

The Wigner entropy enjoys various reasonable properties; in particular it is invariant over all symplectic transformations in phase space or equivalently all Gaussian unitaries in state space. Its excess with respect to  $\ln \pi +1$ is an asymptotic measure of the number of random bits that are needed to generate a sample of the Wigner function from the vacuum state. More generally, since the Wigner entropy is the Shannon differential entropy of the Wigner function, viewed as a genuine probability distribution, it inherits all its key features. For example, we may easily extend to Wigner entropies the celebrated entropy power inequality \cite{Cover1991}, which relates to the entropy of the convolution of probability distributions. Consider the setup of Fig. \ref{fig:bbs_schema} where the input state is again a product state $\hat{\rho}_A\otimes\hat{\rho}_B$ but the beamsplitter now has an arbitrary transmittance $\eta$, so that the output state reads 
\begin{equation}
	\hat{\sigma} = 
	\mathrm{Tr}_{2}
	\left[
	\hat{U}_{\eta}
	\left(
	\hat{\rho}_A\otimes\hat{\rho}_B
	\right)
	\hat{U}_{\eta}^\dagger
	\right].
\end{equation}
Let us restrict to the special case where both $\hat{\rho}_A$ and $\hat{\rho}_B$ are Wigner-positive states, which of course implies that $\hat{\rho}_A\otimes\hat{\rho}_B$ is Wigner positive as well as ${\hat\sigma}$ (even if $\eta\ne 1/2$). Thus,  $\hat{\rho}_A$, and $\hat{\rho}_B$, and ${\hat\sigma}$ all have a well-defined Wigner entropy, which we denote respectively as $h_A$, $h_B$, and $h_{\textrm{out}}$. Since the beam splitter effects the affine transformation $x_{\textrm{out}} = \sqrt{\eta} \, x_A + \sqrt{1-\eta}\,  x_B$ and  $p_{\textrm{out}} = \sqrt{\eta} \, p_A + \sqrt{1-\eta}\,  p_B$ in phase space, we may simply treat this as a convolution formula for probability distributions. Hence, the entropy power inequality directly applies to the Wigner entropy. Defining the Wigner entropy-power \footnote{We define the Wigner entropy-power $N$ of a (Wigner-positive) state as the entropy power of the Wigner function of the state. Here, the entropy power is defined following Shannon information theory for a \textit{pair} of continuous variables $x$ and $p$, namely $N=(2\pi e)^{-1}\textrm{e}^{h(x,p)}$, where $h$ stands for the Shannon differential entropy. } of the two input states as
\begin{equation}
N_A=(2\pi e)^{-1} \textrm{e}^{h_A}, \quad  N_B=(2\pi e)^{-1} \textrm{e}^{h_B},
\end{equation}
and the Wigner entropy-power of the output state as 
\begin{equation}
N_{\textrm{out}} =(2\pi e)^{-1} \textrm{e}^{h_{\textrm{out}} },
\end{equation}
we obtain the \textit{Wigner entropy-power inequality}
\begin{equation}
N_{\textrm{out}} \ge \eta \, N_A + (1-\eta) N_B  .
\end{equation}
This is equivalent to a nontrivial lower bound on the Wigner entropy of the output state $\hat{\sigma}$, namely, $h(W_{\hat{\sigma}}) \ge h(W_{\hat{\sigma}_G})$, where $\hat{\sigma}_G$  denotes the Gaussian output state obtained if each input state is replaced by the phase-invariant Gaussian state (i.e., thermal states) with the same Wigner entropy. This illustrates the physical significance of the Wigner entropy.

Defining the Wigner entropy for Wigner-positive states might also be a good starting point for investigating the states that are \textit{not} within $\mathcal{Q}_+$ and whose Wigner function admits a negative region, hence indicating their nonclassicality and potential computational advantage (we recall that Wigner-positive states are efficiently simulatable classically \cite{Eisert2012}). Just as the characterization of separable states helps understand the advantage offered by entanglement and leads to a resource theory of entanglement, we may envisage building a resource theory of Wigner negativity based on Wigner entropies along the lines of the resource theory of quantum non-Gaussianity \cite{Zhuang2018,Albarelli2018}, going beyond witnesses of Wigner negativity \cite{Chabaud2021}.

Finally, a natural extension of the present work is to consider more than a single harmonic oscillator (or bosonic mode) as we expect that all properties of the Wigner entropy and especially conjecture \eqref{eq:wig_conj} will generalize.  Further, following the lines of a recent work \cite{Floerchinger2021}, we might investigate the detection of entanglement in continuous-variable states by defining a Wigner conditional entropy and Wigner mutual information. Let us mention that this work is part of a broader project. The key observation is that the Wigner function of Wigner-positive states can be interpreted as a true probability distribution. Hence, we can take advantage of this observation and adapt all standard tools of probability theory (here, we have applied Shannon information theory to define the Wigner entropy). In this context, the theory of majorization \cite{Marshall2011} has proved to be another powerful tool, and it notably allows to formulate a generalization of Wehrl conjecture \cite{Lieb2002}. In a forthcoming paper \cite{VanHerstraeten2021Major}, we use the theory of majorization to state a stronger conjecture on the uncertainty content of Wigner functions. This enables us, for instance, to demonstrate analytically the lower bound on $h(W)$ for all phase-invariant Wigner-positive states in $\mathbb{S}^{2}_{+}$, including the dark blue region.

\subsection*{Note added}
We have learned that our method for generating positive Wigner functions with a 50:50 beam splitter as explained in Appendix \ref{apd:bbs_wigner_positive} has recently also been described in \cite{Becker2019}.

\subsection*{Acknowledgments}
The authors warmly thank Christos Gagatsos, Anaelle Hertz, Michael G. Jabbour and Karol \.Zyczkowski for helpful discussions on this subject. Z.V.H. acknowledges a fellowship from the FRIA foundation (F.R.S.-FNRS). N.J.C. acknowledges support by the F.R.S.-FNRS under Project No. T.0224.18 and by the EC under project ShoQC within ERA-NET Cofund in Quantum Technologies (QuantERA) program.

\appendix

\section{Wigner-R\'enyi entropy}
\label{sect-wigner-renyi}

The Shannon differential entropy is an uncertainty measure that belongs to a broader family, known as R\'enyi differential entropies.
Just as we defined the Wigner entropy of a Wigner-positive state as the Shannon differential entropy of its Wigner function, it is natural to define the Wigner-R\'enyi entropy of a Wigner-positive state as
\begin{equation}
	h_{\alpha}\left(W\right)
	=
	\frac{1}{1-\alpha}
	\ln\left(
	\iint
	[ W(x,p) ]^\alpha
	\, \mathrm{d}x\, \mathrm{d}p
	\right),
\end{equation}
where $\alpha\ne 1$ is a real non-negative parameter.

Interestingly, some values of $\alpha$ are endowed with a special meaning. Denoting as $\mathrm{supp}(W)$ the part of the domain of $W$ where $W>0$ and denoting  as $\nu$ the Lebesgue measure, we have $h_0(W)=\ln(\nu[\mathrm{supp}(W)])$ when the parameter $\alpha=0$. This diverges when applied to any Wigner function $W$ since the size of the support of $W$ is infinite. In the limit $\alpha\rightarrow1$, $h_\alpha$ tends to the Shannon differential entropy, so that $h_1(W)$ coincides with $h(W)$.
The R\' enyi entropy with parameter $\alpha=2$ is sometime called the collision entropy, and, applied to a Wigner function $W$, it is related to the purity of the corresponding state. 
Denoting the purity as $\mu=\mathrm{Tr}\left[\hat{\rho}^2\right]=2\pi\iint [W(x,p)]^2\, \mathrm{d}x\, \mathrm{d}p$, we have the relation $h_2(W)=\ln(2\pi/\mu)$.
Finally, the case $\alpha\rightarrow\infty$ can be related to the maximum value of $W$ as $h_\infty(W)=-\ln [ \max_{x,p} W(x,p) ]$.

Note that, following the same reasoning as in Sec.~\ref{sec:wig_entropy}, we observe that the Wigner-R\'enyi entropy is invariant under symplectic transformations in phase space (i.e., Gaussian unitaries in state space). The Wigner-R\'enyi entropy of the vacuum state (or any pure Gaussian state) gives
\begin{equation}
	h_{\alpha}(W_0)
	=
	\ln\pi
	+
	\frac{\ln\alpha}{\alpha-1}.
\end{equation}
Then, in the same spirit as conjecture \eqref{eq:wig_conj}, we conjecture that the Wigner-R\'enyi entropy of any Wigner-positive state is lower bounded by the value it takes for the vacuum:
\begin{equation}
h_{\alpha}\left(W_{\! \hat{\rho}}\right) \geq h_{\alpha}\left(W_0\right)\qquad  \forall \hat{\rho}\in \mathcal{Q}_+
\end{equation}
Of course, it coincides with conjecture \eqref{eq:wig_conj} when $\alpha\rightarrow1$.
Let us examine this Wigner-Rényi conjecture for other special values of the parameter $\alpha$, namely,
\begin{align}
	h_2(W)&\geq \ln 2\pi ,
	\label{eq:h_alpha_2}
	\\[0.5em]
	h_\infty(W)&\geq \ln\pi
	\label{eq:h_alpha_inf}.
\end{align}
For $\alpha=2$, the fact that the purity $\mu$ is upper bounded by $1$ implies Eq.~\eqref{eq:h_alpha_2}.
Also, the Wigner function of any state is upper bounded by $1/\pi$, which implies Eq.~\eqref{eq:h_alpha_inf} for $\alpha\to\infty$. Furthermore, for $\alpha= 0$, the Wigner-R\'enyi conjecture implies that the support of any Wigner function is unbounded, which is a well-known fact. These elements support the validity of the Wigner-R\'enyi conjecture and especially conjecture \eqref{eq:wig_conj} when $\alpha\rightarrow1$.

\section{A balanced beam-splitter produces Wigner-positive states}
\label{apd:bbs_wigner_positive}

In this Appendix, we show that when a balanced beam splitter is fed by a two-mode separable input, then its reduced single-mode output is Wigner positive.
We consider the following setup:
\begin{equation}
	\hat{\sigma} = \mathrm{Tr}_B
	\left[
	\hat{U}_{\frac{1}{2}}
	\left(
	\hat{\rho}_A
	\otimes
	\hat{\rho}_B
	\right)
	\hat{U}_{\frac{1}{2}}^{\dagger}
	\right],
\end{equation}
where $\hat{U}_{\eta}$ is the unitary operator of the beam splitter,
\begin{equation}
	\hat{U}_\eta =
	\exp\left(
	\theta
	\left(
	\hat{a}^\dagger\hat{b}
	-
	\hat{a}\hat{b}^\dagger
	\right)
	\right),
\end{equation}
such that $\eta=\cos^2\theta$ and $0\leq\theta\leq\pi/2$.
We are going to show that $\hat{\sigma}$ is Wigner positive when $\eta = 1/2$:
\begin{equation}
	W_{\hat{\sigma}}(x,p)\geq 0
	\qquad\forall x,p.
\end{equation}
Our proof has two parts. 
In the first part we show that the action of a beam splitter in phase space corresponds to a convolution between the Wigner functions of the two inputs.
In the second part, we show that the convolution of two Wigner functions corresponds in state space to the overlap between two density operators, which is always a non-negative quantity.
Finally, we give some further observations regarding the $\hat{\sigma}(m,n)$ states which are built using that setup.

\subsection{Convolution in phase space}

The action of a beam splitter on the mode operators is described by the following transformation:
\begin{equation}
	\begin{pmatrix}
		\hat{a}'
		\\
		\hat{b}'
	\end{pmatrix}
	=
	\begin{pmatrix}
		\sqrt{\eta} &\sqrt{1-\eta}
		\\
		-\sqrt{\eta} &\sqrt{\eta}
	\end{pmatrix}
	\begin{pmatrix}
		\hat{a}
		\\
		\hat{b}
	\end{pmatrix}.
\end{equation}
From this, we derive the expression of the old quadrature operators $\hat{x}_A$, $\hat{x}_B$, $\hat{p}_A$, and $\hat{p}_B$ as a function of the new quadrature operators $\hat{x}_A'$, $\hat{x}_B'$, $\hat{p}_A'$, and $\hat{p}_B'$:
\begin{equation}
	\begin{cases}
		\hat{x}_A
		&=
		\sqrt{\eta}
		\ \hat{x}_A'-\sqrt{1-\eta}\ \hat{x}_B',
		\\[0.5em]
		\hat{p}_A
		&=
		\sqrt{\eta}\ \hat{p}_A'-\sqrt{1-\eta}\ \hat{p}_B',
		\\[0.5em]
		\hat{x}_B
		&=
		\sqrt{1-\eta}\ \hat{x}_A'+\sqrt{\eta}\ \hat{x}_B',
		\\[0.5em]
		\hat{p}_B
		&=
		\sqrt{1-\eta}\ \hat{p}_A'+\sqrt{\eta}\ \hat{p}_B'.
	\end{cases}
\end{equation}
The two-mode Wigner function of the output $W'$ is then obtained through the following relation:
\begin{equation}
	\begin{split}
		W'(x_A',p_A',x_B',p_B') &= W(x_A,p_A, x_B, p_B)
		\\
		&= W_A(x_A,p_A)W_B(x_B,p_B),
	\end{split}
\end{equation}
where $W_A$ and $W_B$ are the Wigner functions of respectively $\hat{\rho}_A$ and $\hat{\rho}_B$.
Computing the Wigner function of $\hat{\sigma}$ is done by integrating $W'(x_A',p_A',x_B',p_B')$ over the variables $x_B'$, $p_B'$:
\begin{equation}
	\begin{split}
		W_{\hat{\sigma}}\left(x_A',p_A'\right)
		=
		\iint\mathrm{d}x_B'\mathrm{d}p_B'
		\\
		\times W_A\left(
		\sqrt{\eta}\ x_A'-\sqrt{1-\eta}\ x_B',
		\sqrt{\eta}\ p_A'-\sqrt{1-\eta}\ p_B'
		\right)
		\\
		\times W_B\left(
		\sqrt{1-\eta}\ x_A'+\sqrt{\eta}\ x_B',
		\sqrt{1-\eta}\ p_A'+\sqrt{\eta}\ p_B'
		\right).
	\end{split}
\end{equation}
That expression finds a natural writing by introducing the new variables $x''$, $p''$:
\begin{equation}
	\begin{cases}
		x'' = \eta\ x_A' - \sqrt{\eta(1-\eta)}x_B',
		\\[1em]
		p'' = \eta\ p_A' - \sqrt{\eta(1-\eta)}p_B',
	\end{cases}
\end{equation}
so that it reduces to
\begin{equation}
	\begin{split}
		W_{\hat{\sigma}}\left(x_A',p_A'\right)
		=
		\iint\mathrm{d}x''\mathrm{d}p''
		\frac{1}{\eta}W_A\left(
		\frac{x''}{\sqrt{\eta}},
		\frac{p''}{\sqrt{\eta}}
		\right)
		\\
		\times
		\hspace{0.5em}
		\frac{1}{1-\eta}W_B\left(
		\frac{x_A'-x''}{\sqrt{1-\eta}},
		\frac{p_A'-x''}{\sqrt{1-\eta}}
		\right).
	\end{split}
	\label{eq:bbs_convolution}
\end{equation}
Equation \eqref{eq:bbs_convolution} shows that the action of a beam splitter on a product input corresponds to a convolution between $W_A$ and $W_B$ rescaled, respectively, by a factor $\sqrt{\eta}$ and $\sqrt{1-\eta}$.
In the case of a balanced beam splitter, the value of the parameter $\eta$ is $1/2$, and the rescaling values are equal. 
Introducing the new variables $\tilde{x}=\sqrt{2}x''$ and $\tilde{p}=\sqrt{2}p''$, we can then write the previous expression as
\begin{equation}
	\begin{split}
		W_{\hat{\sigma}}\left(x_A',p_A'\right)
		=
		2
		\iint\mathrm{d}\tilde{x}\mathrm{d}\tilde{p}
		W_A\left(
		\tilde{x},
		\tilde{p}
		\right)
		\\
		\times
		\hspace{0.5em}
		W_B\left(
		\sqrt{2}x_A'-\tilde{x},
		\sqrt{2}p_A'-\tilde{p}
		\right).
	\end{split}
	\label{eq:bbs_balanced}
\end{equation}

\subsection{State space picture}

Equation \eqref{eq:bbs_balanced} can be expressed in state space formalism.
To that purpose, we recall the usual rotation and displacement operators acting on mode $\hat{b}$:
\begin{equation}
	\begin{split}
		\hat{R}(\varphi)
		&=
		\exp(-i\varphi\hat{b}^\dagger\hat{b}),
		\\
		\hat{D}(\alpha)
		&=
		\exp\left(
		\alpha\hat{b}^\dagger-\alpha^\ast\hat{b}
		\right).
	\end{split}
\end{equation}
Let us define $\hat{\rho}^\prime_B$ as the result of a particular combination of $\hat{R}$ and $\hat{D}$ acting on $\hat{\rho}_B$:
\begin{equation}
	\hat{\rho}^\prime_B
	=
	\hat{D}(x_\alpha+ip_\alpha)
	\hat{R}(\pi)
	\hat{\rho}_B
	\hat{R}^\dagger(\pi)
	\hat{D}^\dagger(x_\alpha+ip_\alpha).
	\label{eq:rho_b_prime}
\end{equation}
The Wigner function of $\hat{\rho}^\prime_B$ can be expressed from the Wigner function of $\hat{\rho}_B$ as follows:
\begin{equation}
	\begin{split}
		\hat{\rho}_B
		\quad&\longleftrightarrow\quad
		W_B(x,p)
		\\
		\hat{\rho}^\prime_B
		\quad&\longleftrightarrow\quad
		W_B(\sqrt{2}x_\alpha-x,\sqrt{2}p_\alpha-p).
	\end{split}
	\label{eq:disp_and_rot_wigner}
\end{equation}
The last ingredient of our proof is the expression of the overlap between two quantum states, which is always non-negative.
That quantity can be expressed equivalently in the formalism of state space or phase space:
\begin{equation}
	\mathrm{Tr}\left[
	\hat{\rho}_1\hat{\rho}_2
	\right]
	=
	2\pi\iint\mathrm{d}x\mathrm{d}p
	W_1(x,p)W_2(x,p)
	\geq 0.
	\label{eq:overlap_quantum_states}
\end{equation}
Combining Eqs. \eqref{eq:bbs_balanced}, \eqref{eq:disp_and_rot_wigner}, and \eqref{eq:overlap_quantum_states} gives us the expression of $W_{\hat{\sigma}}$ as an overlap between two quantum states:
\begin{equation}
	W_{\hat{\sigma}}(x_A',p_A')
	=
	\frac{1}{\pi}
	\mathrm{Tr}\left[
	\hat{\rho}_A
	\hat{\rho}^\prime_B
	\right]
	\geq 0,
	\label{eq:sigma_disp_rot}
\end{equation}
where the $x_A',p_A'$ dependence is hidden in $\hat{\rho}_B^\prime$.
This concludes our proof, and we have shown that $\hat{\sigma}$ is Wigner positive.
The proof naturally expands from product inputs to separable inputs since the mixing of Wigner-positive states remains Wigner positive.

\subsection{Beam-splitter $\hat{\sigma}(m,n)$ states}

Beam-splitter $\hat{\sigma}(m,n)$ states are defined in Sec. \ref{sec:wig_pos} and play a particular role in this paper.
We give here the expression of their density operator decomposed onto the basis of Fock states:
\begin{equation}
	\begin{split}
		\hat{\sigma}(m,n)=
		\left(m!n!2^m 2^n\right)^{-1}
		\sum\limits_{z=0}^{m+n}
		\sum\limits_{i=\max(0,z-n)}^{\min(z,m)}
		\sum\limits_{j=\max(0,z-n)}^{\min(z,m)}
		\\
		\binom{m}{i}
		\binom{n}{z-i}
		\binom{m}{j}
		\binom{n}{z-j}
		(-1)^{i+j}
		z!(m+n-z)!
		\ket{z}\bra{z},
	\end{split}
\end{equation}
where $\ket{z}\bra{z}$ is a projector onto the $z^\text{th}$ Fock state.

Taking advantage of Eqs. \eqref{eq:rho_b_prime} and \eqref{eq:sigma_disp_rot}, we can easily show that the Wigner function of $\hat{\sigma}(m,n)$ (that we write $S_{(m,n)}$) cancels at the origin when $m\neq n$.
Indeed, since Fock states are invariant under rotation, it follows that
\begin{equation}
	S_{(m,n)}(0,0)
	=
	\frac{1}{\pi}
	\mathrm{Tr}\left[
	\ket{m}\bra{m}
	\ket{n}\bra{n}
	\right]
	=
	\frac{1}{\pi}
	\delta_{mn},
\end{equation}
where $\delta_{mn}$ is the Kronecker delta.

\section{Wigner-positivity conditions for phase-invariant mixtures up to 2 photons}
\label{apd:mixture_2phot}

In this Appendix, we derive the conditions that apply to a mixture of the first three Fock states (0, 1, and 2) such that it has a non-negative Wigner function.
We are considering any state that can be written as
\begin{equation}
	\hat{\rho} = \left(1-p_1-p_2\right)
	\ket{0}\bra{0}
	+p_1\ket{1}\bra{1}
	+p_2\ket{2}\bra{2},
\end{equation}
where $p_1,p_2\geq 0$ and $p_1+p_2\leq 1$.
We first identify a condition on $p_1$ and $p_2$ that is equivalent to the Wigner positivity of $\hat{\rho}$.
Then we compute the geometrical locus of points in the $(p_1, p_2)$ plane that ensures Wigner positivity, and extremal Wigner positivity.

\subsection{Equivalent condition of Wigner-positivity}

The Wigner function of a Fock state is radial and reads as
\begin{equation}
	W_n(r) = \frac{1}{\pi}(-1)^n\exp(-r^2)L_n(2r^2),
\end{equation}
where we use the non-negative radial parameter $r = \sqrt{x^2+p^2}$.
We recall the three first Laguerre polynomials:
\begin{equation}
	\begin{split}
		L_0(x) &= 1,
		\\
		L_1(x) &= -x+1,
		\\
		L_2(x) &= \frac{1}{2}x^2-2x+1.
	\end{split}
\end{equation}
Using this, we can express the Wigner function of $\hat{\rho}$ as
\begin{equation}
	W(r) = \dfrac{1}{\pi}\exp(-r^2)
	\left(
	2p_2 r^4
	+ \left(2p_1-4p_2\right)r^2
	+1-2p_1
	\right).
	\label{eq:wig_radial_p1_p2}
\end{equation}
We want to identify the set of possible values of $p_1, p_2$ such that $\hat{\rho}$ is Wigner positive.
Introducing the parameter $t=2r^2$, we can write an equivalent condition to $W(r)\geq 0$ $\forall r\geq 0$ as
\begin{equation}
	\frac{1}{2}p_2 t^2
	+(p_1-2p_2)t
	+1-2p_1\geq 0
	\qquad\forall t\geq 0.
	\label{eq:wigner_positivity_parabola}
\end{equation}

\subsection{Locus of Wigner-positivity in the $(p_1, p_2)$ plane}

Equation \eqref{eq:wigner_positivity_parabola} is a second-order polynomial with a non-negative coefficient associated to $t^2$. We want to have non-negative values for all $t\geq 0$.
This is possible either if its discriminant $\Delta$ is nonpositive ($\Delta\leq 0$), or if both its roots correspond to $t\leq 0$.

\begin{figure}
	\includegraphics[width=7cm]{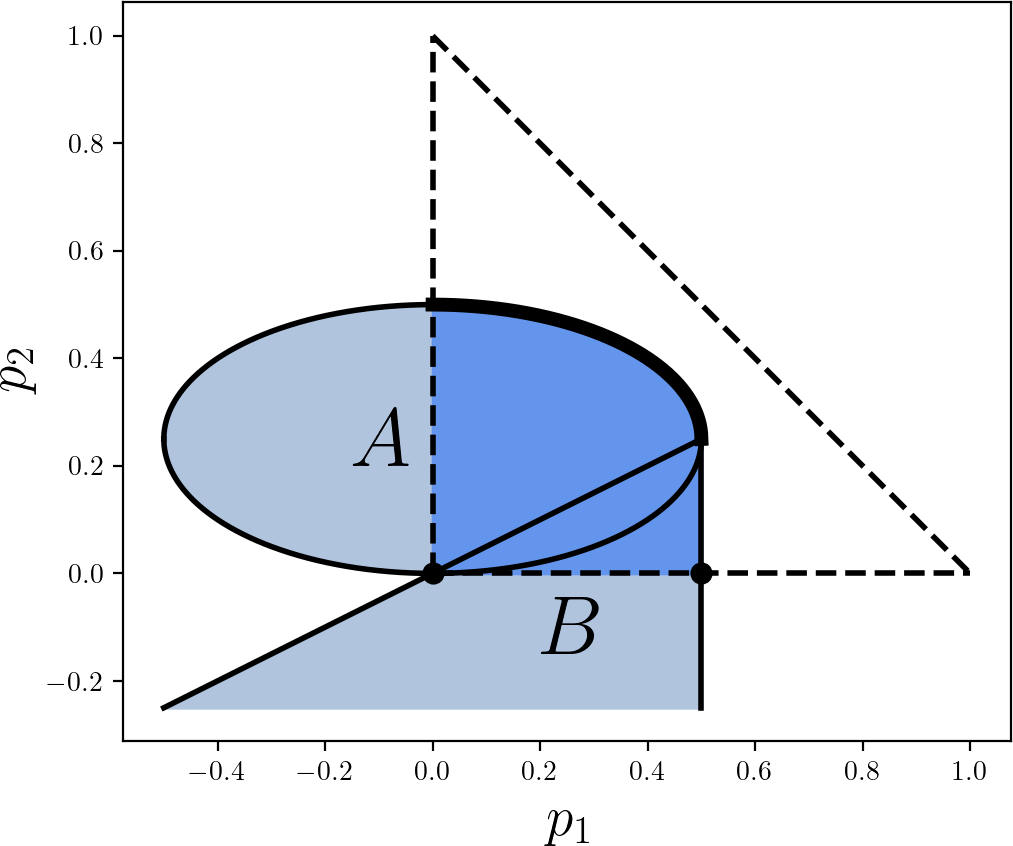}
	\caption{Geometrical locus of Wigner-positivity within the $(p_1,p_2)$ plane, corresponding to the boundary of the dark blue region $\mathbb{S}^{2}_{+}$ satisfying Eq. \eqref{eq:condition_union}. The boundary points and curve of $\mathbb{S}^{2}_{+}$ that are extremal are shown in bold. 
		The ellipse ($A$) corresponds to the region where $(p_1, p_2)$ is such that Eq. \eqref{eq:wigner_positivity_parabola} is never negative.
		The semi-infinite triangular region ($B$) corresponds to values of $(p_1, p_2)$ such that Eq. \eqref{eq:wigner_positivity_parabola} becomes negative only for negative values of $t$.
		The dashed lines forming a triangle define the physicality limits, that is, $p_1, p_2\geq 0$ and $p_1+p_2\leq 1$. The union of ($A$) and ($B$) that belongs to the physicality triangle yields the dark blue region $\mathbb{S}^{2}_{+}$.   }
	\label{fig:geometrical_locus}
\end{figure}

Let us examine the latter possibility first. For a second-order polynomial equation defined by $at^2+bt+c=0$, the sum of its roots is $-b/a$ and their product is $c/a$.
The two roots are nonpositive if their sum is nonpositive and their product is non-negative.
Applied to Eq. \eqref{eq:wigner_positivity_parabola}, this gives the following conditions:
\begin{equation}
	\begin{cases}
		p_1\geq 2p_2
		\\[1em]
		p_1\leq \frac{1}{2}.
	\end{cases}
	\label{eq:condition_portion_plane}
\end{equation}
Condition \eqref{eq:condition_portion_plane} describes a locus which is the intersection of two half planes.
We now check the discriminant condition. 
The discriminant is equal to $\Delta = 4p_2^2-2p_2+p_1^2$, so that the condition $\Delta\leq 0$ can be written as
\begin{equation}
	\left(
	\frac{p_1}{1/2}
	\right)^2
	+
	\left(
	\frac{p_2-1/4}{1/4}
	\right)^2
	\leq
	1.
	\label{eq:condition_ellipse}
\end{equation}
Condition \eqref{eq:condition_ellipse} describes an ellipse.
Note that the union of the sets determined by conditions \eqref{eq:condition_portion_plane} and \eqref{eq:condition_ellipse} alongside with the physicality conditions can be summarized as
\begin{equation}
	\begin{cases}
		p_1\leq\frac{1}{2}
		\\[1em]
		p_2\leq\frac{1}{4}+\frac{1}{4}\sqrt{1-4p_1^2},
	\end{cases}
	\label{eq:condition_union}
\end{equation}
with the additional constraint that $p_1, p_2\geq 0$.
Figure \ref{fig:geometrical_locus} illustrates the geometrical locus associated to the different conditions.

\subsection{Locus of extremal Wigner-positive states}

Let us define $P(t)$ as the second-order polynomial described in Eq. \eqref{eq:wigner_positivity_parabola}.
We refer to the first-order derivative of $P(t)$ with respect to $t$ as $P'(t)$:
\begin{equation}
	\begin{split}
		P(t) &=
		\frac{1}{2}p_2 t^2
		+(p_1-2p_2)t
		+1-2p_1,
		\\[1em]
		P'(t) &= p_2 t + p_1-2p_2.
	\end{split}
\end{equation}
The locus of extremal Wigner-positive states is the set of $(p_1,p_2)$ such that
\begin{equation}
	\exists t\geq 0\quad
	\text{such that}\quad
	\begin{cases}
		P(t) = 0
		\\
		P'(t) = 0.
	\end{cases}
\end{equation}
The condition $P'(t)=0$ is satisfied at $t=2-p_1/p_2$. 
Injecting that value of $t$ in $P(t)=0$ gives us the following equation:
\begin{equation}
	4p_2^2-2p_2+p_1^2=0.
\end{equation}
with the additional constraint that $2p_2\geq p_1$, since $t\geq 0$.
This describes an arc of an ellipse, which we can parametrize as follows:
\begin{equation}
	\begin{cases}
		p_1 = \dfrac{1}{2}\sqrt{1-a^2}
		\\[1em]
		p_2 = \dfrac{1}{4}(a+1),
	\end{cases}
\end{equation}
where the parameter $a$ goes from $0$ to $1$.
Injecting that parametrization in Eq. \eqref{eq:wig_radial_p1_p2} yields the following expression:
\begin{equation}
	W_a(r)=
	\dfrac{1}{\pi}\exp\left(-r^2\right)
	\dfrac{1}{2}(a+1)
	\left(r^2-1+\sqrt{\dfrac{1-a}{1+a}}\right)^2.
\end{equation}
$W_a(r)$ is the radial Wigner function of the extremal Wigner-positive states located on the arc of the ellipse appearing in bold in Fig. \ref{fig:geometrical_locus}.

\section{Convex decomposition of extremal passive states into positive functions}

\label{apd:formula_extremal_states}

The Wigner positivity of extremal passive states, and by extension of the whole set of passive states, is often taken as a known fact \cite{Bastiaans1983}.
However, what is less known is that is in fact only the weakening of a stronger mathematical relationship.
Indeed, going back to the origin of this result, we find the following identity in \cite{Szeg1939}:
\begin{equation}
	\begin{split}
		2^n n!
		\sum\limits_{k=0}^{n}
		(-1)^k
		L_k(2x^2+2y^2)
		\qquad\qquad\qquad
		\\
		\qquad\qquad\qquad
		=
		\sum\limits_{k=0}^{n}
		\begin{pmatrix}
			n\\ k
		\end{pmatrix}
		\left[
			H_k(x)
		\right]^2
		\left[
			H_{n-k}(y)
		\right]^2,
	\end{split}
\label{eq:relation_laguerre_hermite}
\end{equation}
where $L_k$ and $H_k$ are respectively the $k^\text{th}$ Laguerre and Hermite polynomials.
From Eq. \eqref{eq:relation_laguerre_hermite}, it readily appears that the left-hand side is non-negative since the right-hand side is a sum of squared functions.
This naturally implies that the Wigner function $E_n$ of extremal passive states $\hat{\varepsilon}_n$ [see  Eqs. \eqref{eq:wigner_function_fock} and \eqref{eq:def_extremal_states}] is non-negative.

However, it is possible to get an equality out of Eq. \eqref{eq:relation_laguerre_hermite}.
Indeed, multiplying both sides by $\exp(-x^2-y^2)$ and rearranging the normalization factors, we can make appear the Wigner function of Fock states [Eq. \eqref{eq:wigner_function_fock}] in the left-hand side and the wave function of Fock states [Eq. \eqref{eq:wave_function_fock}] in the right-hand side.
Formulated in these terms, Eq. \eqref{eq:relation_laguerre_hermite} becomes then Eq. \eqref{eq:extremal_states_formula}.

\hspace{10cm}

\bibliography{wigent-resub}

\end{document}